\documentclass{vgtc}
\graphicspath{{figures/}{pictures/}{images/}{./}} 

\usepackage{times}                     

\usepackage{tabu}                      
\usepackage{booktabs}                  
\usepackage{lipsum}                    
\usepackage{mwe}                       
\usepackage{endnotes}

\usepackage{wrapfig}
\usepackage{graphicx} 
\usepackage{caption}
\newcommand{\revision}[1]{\textcolor{black}{#1}}

\usepackage{mathptmx}                  
\usepackage{titlesec}
\usepackage{enumitem}
\setlist{itemsep=0.02em}
\setlist[itemize]{left=0pt}
\titleformat{\subsubsection}[runin] 
    {\normalfont\normalsize\bfseries} 
    {\thesubsubsection} 
    {1em} 
    {} 
    [] 

\titlespacing{\subsubsection} 
    {0pt} 
    {2ex plus 1ex minus .2ex} 
    {10pt} 

\onlineid{1075}

\vgtccategory{Research}

\vgtcinsertpkg

\title{MapCraft: Dissecting and Designing Custom Geo-Infographics}

\author{%
Xinyuan Zhang\thanks{Email: Xinyuan.Zhang20@student.xjtlu.edu.cn.}
   \and Yifan Xu \thanks{Email: Yifan.xu2202@student.xjtlu.edu.cn}%
  \and Kaiwen Li\thanks{Email: kaiwen.li18@alumni.xjtlu.edu.cn}%
  \and Lingyun Yu\thanks{Email: Lingyun.Yu@xjtlu.edu.cn}%
  \and Yu Liu\thanks{Email: Yu.Liu02@xjtlu.edu.cn}%
} %
\affiliation{\scriptsize School of Advanced Technology \\ Xi'an Jiaotong-Liverpool University}

\teaser{
  \centering
  \includegraphics[width=\linewidth]{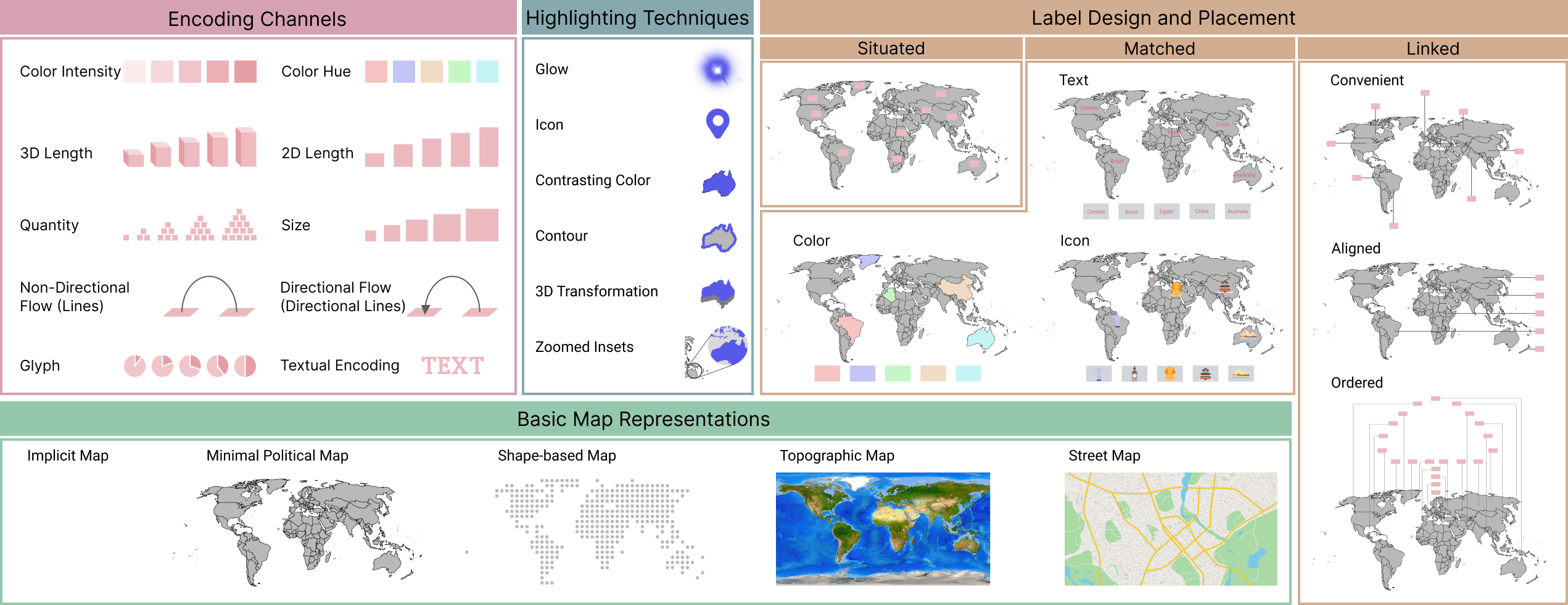}
  \caption{A design space for geo-infographics \revision{ (Note: An Implicit Map means the map itself is not visible, but data points are positioned based on map coordinates).}}
  \label{fig:teaser}
}

\abstract{

Geographic infographics are increasingly utilized across various domains to convey spatially relevant information effectively. However, creating these infographics typically requires substantial expertise in design and visualization, as well as proficiency with specialized tools, which can deter many potential creators. To address this barrier, our research analyzed and categorized 118 geographic infographics and sketches designed by 8 experts, leading to the development of a structured design space encompassing four critical dimensions: basic map representations, encoding channels, label design and placement, and highlighting techniques.
Based on this design space, we developed a web-based authoring tool that allows users to explore and apply these dimensions interactively. The tool's effectiveness was evaluated through a user study involving 12 participants without prior design experience. Participants were first required manually to create geographic infographics using provided datasets, then utilize our authoring tool to recreate and refine their initial drafts.
We also conducted pre- and post-use assessments of the participants' knowledge of geographic infographic design. The findings revealed significant improvements in understanding and applying information encoding channels, highlighting techniques, and label design and placement strategies.
These results demonstrate the tool's dual capacity to assist users in creating geographics while educating them on key visualization strategies. Our tool, therefore, empowers a broader audience, including those with limited design and visualization backgrounds, to effectively create and utilize geo-infographics.

}

\begin{document}

\firstsection{Introduction}

\maketitle

Infographics are a form of static visualization that transforms complex and voluminous data into clear, visually appealing representations, making intricate information accessible and engaging for a broad audience \revision{\cite{Naparin2017}}. By combining visual elements with minimal text, infographics allow audiences to grasp the purpose quickly, focus on key points, and obtain an overview of the data \cite{harrison2015infographic}. This method of visual communication is particularly effective across different cultural and linguistic backgrounds, enhancing comprehension and retention of information \cite{mansour2022use, lerman2021evaluation}. The ability of infographics to support the creation of logical and captivating narratives makes them ideal for concisely presenting complex ideas \cite{10.1145/3377325.3377517, ZHU202024}. Infographics are widely used in various fields, such as education, where they improve learning outcomes by simplifying complex concepts \cite{BRAIN733, gebre2016developing}; business, where they present data-driven insights \cite{young2014infographics}; and healthcare, where they disseminate critical health information to enhance public health literacy \cite{MARTIN201948}.

One specific and significant type of infographic is the geographic infographic, or geo-infographic, which combines traditional thematic mapping with infographic techniques to visually represent spatially relevant information in an engaging and understandable manner\revision{ \cite{he_visualize_2011}}. Geo-infographics have emerged as crucial tools in various fields.
For example, Wang et al. \cite{ijerph19159634} utilized geo-infographics to express landscape ecological risk (LER), aiding environmental scientists in identifying high-risk areas. Similarly, Li et al. \cite{land12061242} employed geo-infographics to display the distribution and changes in land use patterns across different topographic gradients, providing valuable insights for land management and planning. Moreover, Dailey et al. \cite{doi:10.1177/09520767221140954} extended the use of geo-infographics to governmental agencies, enabling effective communication of planning, evaluation, and risk-related information to the broader public.

Despite the growing prevalence of geo-infographics, creating them still requires a high level of expertise in visualization techniques and proficiency with advanced design tools like Adobe Illustrator or Photoshop. This necessity for specialized skills creates a barrier for many potential users. Moreover, existing commercial tools for map visualization, \revision{for example, Google Data Studio, Google Map, and Tableau}  often lack the flexibility needed for creating highly customized geo-infographics, offering only a limited selection of visualization options.

To address these challenges, we propose a comprehensive design space for geo-infographics, which informs the development of geo-infographic authoring tools, \textit{MapCraft}. We conducted a user study with 12 participants to evaluate its usability and educational impact.
Our contributions can be summarized as follows: 
\begin{enumerate}[left=0pt]
    \item[(1)] We decompose the design elements of geo-infographics into a comprehensive design space (see \cref{fig:teaser}).
    \item[(2)] We develop a web-based authoring tool, \href{https://mapcraftforgeo-infographics.github.io/geo-infographics/}{\textit{MapCraft}}, based on this design space, enabling more people to create geo-infographics.
    \item[(3)] We offer findings and reflections based on results from user studies, which provide suggestions on future geographical data visualization and visualization authoring and education tools.
\end{enumerate}

\section{Related Work}
\subsection{Geographical Visualization}
Geographic information has always been essential for understanding and navigating our world \cite{Wieczorek2009}. From ancient times, when maps were etched into the ground, to the advent of precise measurement tools and data collection technologies, the evolution of geographic visualization has been remarkable. Early maps provided basic spatial understanding, but modern techniques have enabled the creation of accurate and detailed maps. 
The development of geographic information systems (GIS) such us ArcGIS \cite{Ormsby2004} and QGIS (Quantum GIS) \cite{Graser2016} allow for the collection, storage, and analysis of spatial data, leading to more sophisticated and informative maps \cite{Wieczorek2009}. 

The ability to layer information onto maps has further enhanced their utility, giving rise to interactive map systems. These systems enable users to explore and manipulate spatial data in real-time. For instance, Chen et al. \cite{VAUD} developed the Visual Analyzer for Urban Data (VAUD), which aggregates and visualizes various types of information—cyber, physical, social, and temporal—on maps, facilitating effective querying and exploration of urban datasets. Zhou et al. \cite{ZHOU2020244} have created visualization interfaces for large-scale geo-tagged social media data, and Bosch et al. \cite{ScatterBlogs2} introduced ScatterBlogs2, a tool for real-time monitoring and analysis of microblog messages, crucial for emergency management and disaster response. Additionally, Splechtna et al. \cite{PB-VRVis-2023-023} implemented an interactive map view allowing users to visualize various statistical values directly on maps, contextualizing spatial relationships to analyze urban economic health. Li et al. \cite{GeoCamera} proposed GeoCamera, which democratizes map visualization by enabling users without filmmaking expertise to design camera movements for geographic data videos. \revision{Lei et al. \cite{Lei2023} developed GeoExplainer to help analysts create explainable documentation for spatial analyses.} These examples underscore the versatility and indispensability of interactive maps in various applications.

While interactive map systems are powerful tools for exploring and analyzing spatial data, there is also a need to present and disseminate this data effectively. This is where geo-infographics come into play. Geo-infographics combine the precision of maps with the visual storytelling power of infographics, presenting spatially relevant information in a static yet visually appealing manner. They are designed to convey complex geographic data clearly and concisely, making them ideal for communication and information dissemination \cite{roth2021cartographic, Song_Roth_Houtman_Prestby_Iverson_Gao_2022}. Unlike interactive maps that require user engagement for data exploration, geo-infographics provide a straightforward, static representation that can be easily shared and understood by a broad audience.

Both interactive maps and geo-infographics are valuable tools, each serving distinct purposes. Interactive maps excel in dynamic data exploration and real-time analysis, catering to users who need to interact with and manipulate spatial data \cite{https://doi.org/10.1111/cgf.14031}. On the other hand, geo-infographics are suited for static presentations that prioritize clarity, visual appeal, and ease of understanding, making them ideal for conveying complex information to a wide audience without requiring user interaction \cite{he_visualize_2011}.

\subsection{Infographics}

Infographics transform complex and voluminous data into clear, visually compelling representations, simplifying intricate information into understandable graphic forms \cite{harrison2015infographic}. They offer numerous advantages, such as enabling audiences to quickly grasp their purpose, focus on key points, and obtain an overview of the data through engaging visual elements \cite{mansour2022use}. The predominant use of graphics with minimal text makes infographics a powerful tool for communication across different cultural and linguistic backgrounds, effectively transcending language barriers \cite{lerman2021evaluation}.

Despite the widespread benefits and applications of infographics, creating high-quality infographics remains a challenging task. This complexity has led to the development of various authoring tools designed to facilitate the creation process. For instance, Wang et al. \cite{DataShot} explored the general infographic design space at both the sheet and element levels and presented Datashot, which enables the automatic generation of fact sheets from tabular data. Additionally, Cui et al. \cite{Text-to-Viz} developed a system capable of automatically generating proportional infographics from natural language statements. Lu et al. \cite{VIF} introduced the concept of Visual Information Flow (VIF), elucidating the semantic structure that connects graphical data elements to convey information and stories, thereby aiding in the creation of better infographic designs and guiding the general visual organization of narratives. Chen et al. \cite{chen2019towards} applied deconstruction and reconstruction techniques to analyze various visual elements within bitmap timeline infographics, aiming to utilize the identified content for the automated generation of new infographics. Similarly, Matthew et al. \cite{TimelinesRevisited} proposed a temporal design space for narrative storytelling, deconstructing timelines into representation, scale, and layout dimensions. These tools aim to simplify the design process, making it more accessible for users with varying levels of expertise.

However, the creation process is still relatively cumbersome when it comes to geo-infographics, which integrate geographic data with infographic techniques. Geo-infographics are valuable for presenting spatially relevant information but require specialized knowledge and skills to design effectively.

\revision{Current tools like Tableau and ManyEyes \cite{ManyEyes}, although powerful, are not specifically designed for geo-infographics and lack comprehensive visualization methods for this purpose. 
While these tools can help users quickly create interactive maps or geo-infographics, they do not fully encompass all possible styles of geo-infographics. As a result, the available variety of geo-infographic styles in these systems is not as wide and comprehensive as what can be found in the market. Moreover, these tools do not list or support all design possibilities, restricting users' creativity and flexibility in creating geo-infographics.}
There is a significant need for dedicated authoring tools tailored specifically for geo-infographics. Such tools should support the detailed examination and categorization of geo-infographic elements, similar to how other infographics have been studied and developed. Drawing on insights from visualization grammars, which use a bottom-up approach starting from data and employing aesthetic mappings to shape the visual form \cite{GraphicsGrammar}, can inform the design of these tools.
Our work addresses this gap by developing \textit{MapCraft}, a geo-infographic authoring tool that integrates both creation and educational functionalities.

\section{Methodology}
Our study is founded on a systematic methodology that incorporates best practices in visualization research and design. This multifaceted strategy began with the compilation of a diverse dataset of 118 geo-infographics sourced from various platforms. Additionally, we involved eight visualization experts who created hand-drawn geo-infographics to supplement design perspectives.
The next phase involved the deconstruction of these geo-infographics to identify fundamental visual elements. This analysis led to the development of a design space encompassing four dimensions: basic map representations, encoding channels, label positions, and highlight techniques. These dimensions are mutually exclusive and provide a comprehensive framework for understanding geo-infographic design.
Utilizing this design space, we developed a web-based authoring tool, \textit{MapCraft}, allowing users to explore and apply these design principles digitally. To ensure the tool's effectiveness and usability, we conducted a user study with 12 participants who had limited or no data visualization experience. This study included usability inquiries and pre- and post-knowledge assessments to measure the educational impact of \textit{MapCraft}.
By adhering to this systematic methodology grounded in established visualization frameworks and user-centered development principles \cite{Wieczorek2009}, we aim to create a geo-infographic authoring tool that is a practical resource for educational and professional use.

\section{Design Space}
\revision{To gain a comprehensive understanding of geo-infographics, we conducted keyword searches across multiple sources, including visualization books \cite{InformationGraphics} and online platforms like Pinterest\footnote{\url{https://www.pinterest.com/}} and Google Images\footnote{\url{https://images.google.com/}}. We used multiple related keywords such as ``geo-infographics'', ``geographic infographics'' and ``map infographics''. From these searches, we selected 118 geo-infographics that were clear, comprehensible, and non-redundant (see examples in \cref{fig:corpus} in Appendix). This diverse and carefully curated dataset forms the basis for our detailed analysis.}

To further refine our understanding and ensure the practical applicability of our findings, we conducted expert reviews with eight experienced visualization researchers. This participant group included 2 females and 6 males, aged between 20 and 30 (average age 24.6, SD=2.7). Six participants were researchers from the Human-Computer Interaction and Visualization Laboratory, including two doctoral students, three postgraduate students, and one undergraduate student. The remaining two participants were undergraduate students participating in multiple data visualization projects. All participants majored in computer-related programs and had experience with at least two visualization tools, including Python, Tableau, D3.js, Excel, and Figma.

The expert review process was structured into four stages, each lasting approximately half an hour. It began with an introduction to the review process and objectives to ensure participants understood the procedures and aims. This was followed by a pre-questionnaire to gather background information about the participants. Subsequently, participants were tasked with drawing two geo-infographics based on provided data: one depicting quantitative data related to population and the other illustrating categorical data concerning country information. We encouraged the experts to enumerate all possible methods they could think of and select what they considered the most suitable and effective approaches. Finally, participants were interviewed to explain their designs and provide suggestions for improvement.

Through this process, we obtained 16 hand-drawn geo-infographics from the experts. Combined with the 118 geo-infographics we gathered from various sources, these formed our comprehensive corpus. We identified common themes and techniques by thoroughly analyzing the visual elements used in the hand-drawn and sourced geo-infographics. This analysis ultimately led to the development of our design space for geo-infographics, organized into four dimensions: basic map representations for geographic information, encoding channels for topical data, label design and placement, and highlighting techniques for important data points (see \cref{fig:teaser}). 

\subsection{Dimension 1: Basic Map Representations}
The basic map representation frequently serves as the foundational component located at the bottom layer of a geo-infographic. It displays geographical location, and all other data matches its corresponding location.

\noindent\textbf{\MakeUppercase{Implicit Map}. }
This category involves geo-infographics that do not explicitly display any visible geographic information elements (see \cref{fig:corpus}.a in Appendix). The geographical context is implicitly present as subsequent data overlays reference geographic locations, but the map lacks borders or country or city labels. This approach allows the audience to focus entirely on the data without any distractions from geographic elements.

\noindent\textbf{\MakeUppercase{Minimal POLITICAL MAP}. }
This type of map representation includes only the most basic geopolitical boundaries. This means that only the country's borders are shown on a world map. A national map might include just the borders between states or provinces. This minimalistic approach ensures that the geographical framework is clear without additional details that could distract from the presented main data.

\noindent\textbf{\MakeUppercase{SHAPE-BASED MAP}. }
Shape-based maps utilize basic shapes, such as dots, squares, or custom icons, to form recognizable boundaries of countries, states, or regions. Unlike implicit maps, which lack visible geographic cues, shape-based maps offer a visual representation of geographic boundaries by arranging these shapes in ways that outline different areas. The shapes themselves do not encode additional data but are purely stylistic elements that help convey the geographic contours of regions.

These maps can be constructed using shapes of uniform size, creating a clear and precise boundary between areas. For instance (see \cref{fig:corpus}.b in Appendix), a country could be depicted using an array of equally sized dots, forming a distinct and recognizable shape that accurately represents its borders. Alternatively, shape-based maps might use shapes of varying sizes to represent regions. In such cases (see \cref{fig:corpus}.c in Appendix), while the overall geographic outline remains discernible, the boundaries between different regions may appear less precise. This approach can introduce a degree of abstraction, making the map visually engaging but potentially less accurate in depicting exact geographic boundaries.

\noindent\noindent\textbf{\MakeUppercase{TOPOGRAPHIC MAP}. }
As the name implies, topographic maps illustrate topographic features like natural elevations and depressions. They provide the audience with a reference to topographic information, facilitating the display of additional details such as land use, climate zones, and natural resources.

\noindent\textbf{\MakeUppercase{Street Map}. }
Street maps concentrate on smaller blocks within cities or countries. They depict urban features such as landmarks, roads, buildings, and traffic information.

\subsection{Dimension 2: Encoding channels}
The encoding channel refers to various visual variables employed to convey focusing information on top of the basic map. We mainly refer to the design space of channels proposed by Tamara \cite{munzner2014visualization}. We considered both quantitative and categorical data. 

Based on the geo-infographics we collected and analyzed, we summarize the most commonly used encoding channels in geo-infographics. 

\setlength\intextsep{0pt}
\setlength\columnsep{0pt}
\begin{wrapfigure}{l}{.12\columnwidth}
    \vspace*{\fill}
    \includegraphics[height=1.8\baselineskip]{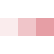}
    \vspace*{\fill}
\end{wrapfigure}
\noindent\textbf{\MakeUppercase{Color (Intensity)}. }
Color intensity refers to the luminance or saturation of a color. This gradient is valuable for visualizing changes in quantitative data size, with low intensity denoting smaller values and high intensity signifying larger ones.

\setlength\intextsep{0pt}
\setlength\columnsep{0pt}
\begin{wrapfigure}{l}{.12\columnwidth}
    \vspace*{\fill}
    \includegraphics[height=1.8\baselineskip]{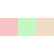}
    \vspace*{\fill}
\end{wrapfigure}
\noindent\textbf{\MakeUppercase{Color (Hue)}. }
Color hue denotes the pure spectrum of red, blue, and green colors. It is commonly employed to represent different categories or groups within the data. For instance, continents may be distinguished by assigning each a distinct hue — blue for Europe, green for Asia, yellow for Africa, red for the Americas, and orange for Oceania.
\revision{Color hue can also represent quantitative data by mapping numeric variable values to distinct hues along a perceptually ordered color palette. For example, 
this gradation from blue to red symbolically illustrates the transition from cooler to warmer temperatures.}

\setlength\intextsep{0pt}
\setlength\columnsep{0pt}
\begin{wrapfigure}{l}{.12\columnwidth}
    \vspace*{\fill}
    \includegraphics[height=1.8\baselineskip]{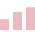}
    \vspace*{\fill}
\end{wrapfigure}
\noindent\textbf{\MakeUppercase{2D Length}. }
2D Length facilitates data representation within the geographical context by overlaying rectangles onto corresponding positions on a map. The height of each line or rectangle is directly proportional to the value of each data point.

\setlength\intextsep{0pt}
\setlength\columnsep{0pt}
\begin{wrapfigure}{l}{.12\columnwidth}
    \vspace*{\fill}
    \includegraphics[height=1.8\baselineskip]{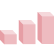}
    \vspace*{\fill}
\end{wrapfigure}
\noindent\textbf{\MakeUppercase{3D Length}. }
3D Length employs three-dimensional rectangular prisms overlaid onto a basic map. Each prism's height correlates directly with the magnitude of the each data point. 

\setlength\intextsep{0pt}
\setlength\columnsep{0pt}
\begin{wrapfigure}{l}{.12\columnwidth}
    \vspace*{\fill}
    \includegraphics[height=1.8\baselineskip]{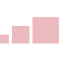}
    \vspace*{\fill}
\end{wrapfigure}
\noindent\textbf{\MakeUppercase{Size}. }
Size represents values through variations in the size of graphical elements such as different shapes, icons, or symbols. For instance, circles can represent retail stores, with their sizes indicating the number of sales at each store \cite{peterson2020gis}. 
Another approach is to distort the geographic area of each region, such as a province or country, in proportion to a specific data value. This category can also be called `` Cartogram '' \cite{gastner2004diffusion, tobler2004thirty}. This method resizes the area to reflect the data magnitude (see \cref{fig:corpus}.d in Appendix), often resulting in no longer geographically accurate shapes that effectively communicate the intended information.

\setlength\intextsep{0pt}
\setlength\columnsep{0pt}
\begin{wrapfigure}{l}{.12\columnwidth}
    \vspace*{\fill}
    \includegraphics[height=1.8\baselineskip]{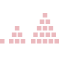}
    \vspace*{\fill}
\end{wrapfigure}
\noindent\textbf{\MakeUppercase{Quantity}. }
Quantity is achieved through the strategic deployment of graphical entities, such as different shapes, icons, or symbols, wherein the quantity of these elements directly corresponds to the underlying numerical values. 
For instance, in the context of demographic mapping, many representative icons or figures might depict a higher population density in a particular region, each symbolizing a specific number of individuals.

\setlength\intextsep{0pt}
\setlength\columnsep{0pt}
\begin{wrapfigure}{l}{.12\columnwidth}
    \vspace*{\fill}
    \includegraphics[height=1.8\baselineskip]{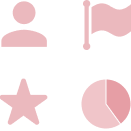}
    \vspace*{\fill}
\end{wrapfigure}
\noindent\textbf{\MakeUppercase{Glyph}. }
Glyphs are versatile tools for visually representing and communicating data through complex icons or composite graphics \cite{Glyph-based}. They can encapsulate multiple quantitative or categorical data points within a visual element. For example, different icons can represent various animals in different countries, or bar charts and pie charts can overlay a map to show demographic information such as age distribution or gender ratios within each country. This approach represents multiple variables or categories within a single glyph, maintaining visual clarity while conveying rich, layered information.

\setlength\intextsep{0pt}
\setlength\columnsep{0pt}
\begin{wrapfigure}{l}{.12\columnwidth}
    \vspace*{\fill}
    \includegraphics[height=1.8\baselineskip]{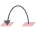}
    \vspace*{\fill}
\end{wrapfigure}
\noindent\textbf{\MakeUppercase{Directional Flow (Directional Lines)}. }

\revision{
Directed flow lines are always used to describe sequential or directed movement or motion, such as migration patterns, trade routes, or flight paths \cite{Zeng2017}.
More detailed examinations of flow lines in geo-infographics can be found in Zhao et al.'s work \cite{FlowLine}. }

\setlength\intextsep{0pt}
\setlength\columnsep{0pt}
\begin{wrapfigure}{l}{.12\columnwidth}
    \vspace*{\fill}
    \includegraphics[height=1.8\baselineskip]{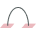}
    \vspace*{\fill}
\end{wrapfigure}
\noindent\textbf{\MakeUppercase{Non-Directional Flow (Lines)}.} Non-directional flow lines represent connections between two places without implying a specific direction. These lines illustrate bi-directional or non-directional relationships, such as diplomatic relations or transportation links between cities.

\setlength\intextsep{0pt}
\setlength\columnsep{0pt}
\begin{wrapfigure}{l}{.12\columnwidth}
    \vspace*{\fill}
    \includegraphics[height=1.8\baselineskip]{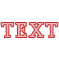}
    \vspace*{\fill}
\end{wrapfigure}
\noindent\textbf{\MakeUppercase{Text}. }
Text involves overlaying text on the map to present information directly. This method is always used when the data is straightforward and concise, allowing designers to exhibit the data through textual means without employing additional visual variables (see \cref{fig:corpus}.a in Appendix).

\begin{figure}[tb]
 \centering
 \includegraphics[width=\columnwidth]{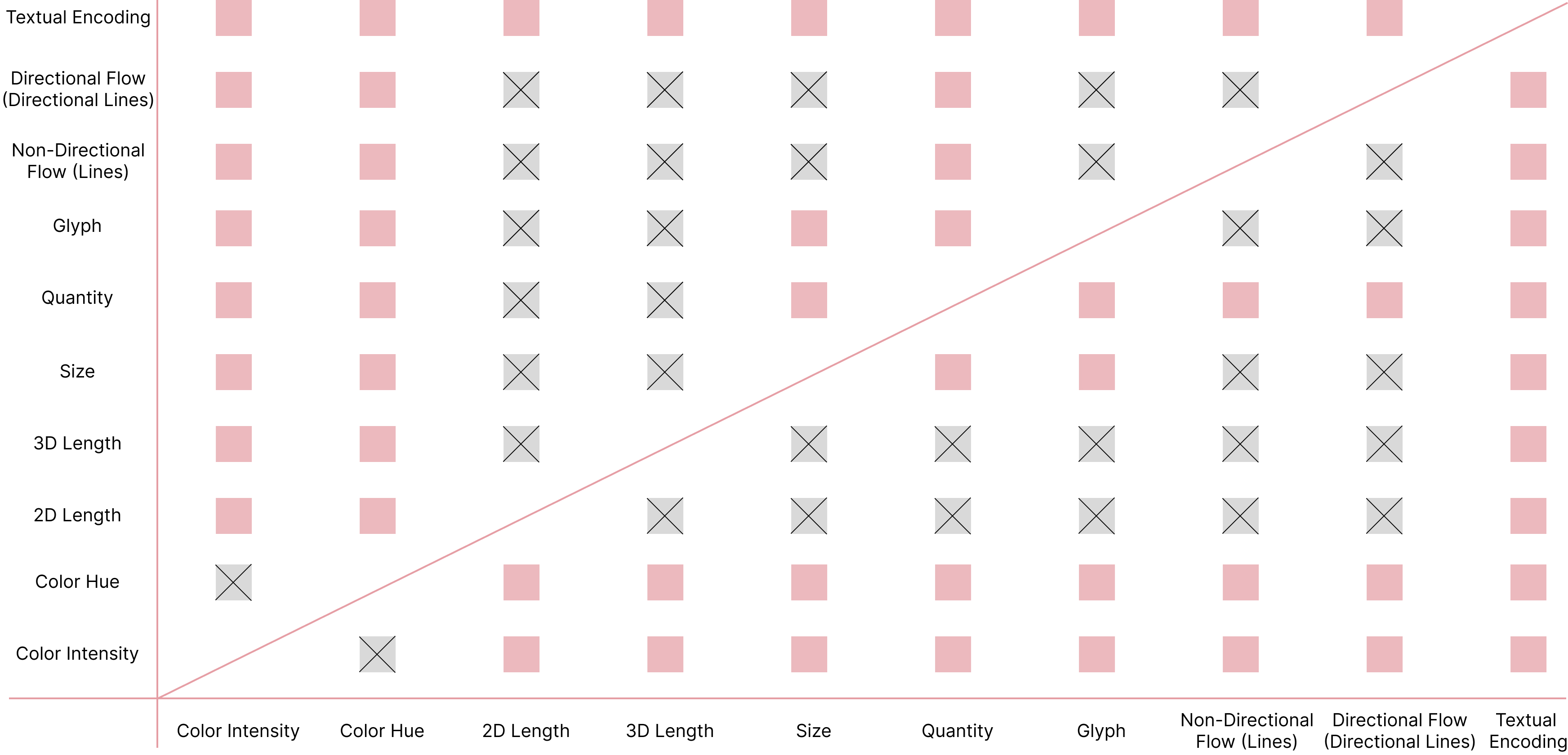}
 \caption{Compatibility Matrix for Dual Encoding Channels. Filled squares indicate effective combinations; crosses (X) denote infeasible combinations.}
 \label{fig:dualEncoding}
\end{figure}

\subsubsection{Dual Encoding Compatibility}
We found that single-channel encodings form the basis of information representation, and many scenarios benefit from combining multiple channels to enhance data comprehension and retention. Recognizing this, we systematically examined the identified single channels and explored their potential pairings. This analysis provides insights and guidelines for effective dual encoding strategies in geo-infographic design. \cref{fig:dualEncoding} details the viable and non-viable combinations for such encoding strategies.

Channels with inherently similar properties cannot be combined. For instance, Color Intensity and Color Hue both encode information using the color spectrum and thus cannot be paired. Similarly, 2D Length and 3D Length fundamentally use the concept of length to represent data, making their combination inappropriate. Directional Flow and Non-Directional Flow, which represent directional relationships through arrows and non-directional relationships through lines, respectively, are mutually exclusive and cannot be integrated. Size and 2D/3D Length cannot be effectively combined due to their similar properties; length can be regarded as the size of a line.

On the other hand, Text can be combined with all other encoding channels. Color is another versatile encoding channel that can be combined with most other channels. Color Intensity and Color Hue can be paired with other encoding channels. Specifically, combining Color Intensity with 2D/3D Length can represent quantitative data, where lower values are depicted with lighter colors and shorter rectangles or cubes and higher values with darker colors and taller rectangles or cubes. The combination of Color Hue with 2D/3D Length operates on the same principle, substituting light colors for lower numerical values and dark colors for higher numerical values. When the data has both category and quantity attributes, Color Hue can be used to distinguish categories, while 2D/3D Length expresses the numerical size. For instance, Color Hue and 2D/3D Length can represent categorical and quantitative GDP attributes, respectively, with Color Hue distinguishing between developed (blue), emerging (green), and developing (orange) nations. The GDP values are depicted through the vertical extent of 2D rectangles or 3D bars, where a taller bar corresponds to a larger GDP and a shorter bar to a smaller GDP. Similarly, Color combined with Size or Quantity follows the same principle as Color with 2D/3D Length.

In specific scenarios, Glyphs cannot be combined with Color. For example, icons of famous scenic spots or national flags possess multiple colors, making overlaying colors on them visually confusing. However, glyphs like triangles or pentagrams representing different data categories can accommodate color coding. For instance, a human-shaped glyph overlaying color encoding enhances the representation of population data. Directional Flow and Non-Directional Flow, representing data flow in geo-infographics, are mutually exclusive and cannot be combined. However, each can independently use a color overlay to encode data type and direction. Glyphs can encode quantitative attributes with varied sizes or indicate data importance through glyph size. While size and quantity can theoretically be superimposed to express quantitative attributes, this may lead to audience confusion, as observed when both size and quantity equate to the same value, potentially misconstrued as multiplication. Quantity can also be used in combination with glyphs to display quantity data. It can also be combined with Directional Flow or Non-Directional Flow to express the magnitude of the circulating data.

\subsection{Dimension 3: Label Design and Placement}
Geo-infographics often need to convey a large amount of complex information. To fully convey the complexity and nuances of the given topic data, it is often necessary to supplement the primary visual encoding with additional annotations in a geo-infographic.
The placement strategy for these labels can be categorized into three distinct approaches: situated, matched, and linked.
\subsubsection{Situated}
Situated labels are superimposed directly onto the map regions they represent. This direct overlay approach ensures a clear and immediate visual connection between the label and the corresponding geographical area, which is particularly useful when the map's regions are sufficiently large to accommodate the text without causing visual clutter.

\subsubsection{Matched}
Matched labels are used when the label information is complex or extensive, and separating the explanatory information from the map is beneficial. This approach uses visual encoding channels to establish clear and intuitive relationships between the map and the accompanying explanatory text. Common methods for creating these matches include color, icons, and text:

\noindent\textbf{\MakeUppercase{Text}. }
This method involves placing concise, descriptive keywords or phrases (such as names of locations or addresses) directly on the map. These keywords act as prompts corresponding to more detailed explanations or annotations in a separate, often blank, infographic section. The reader can easily connect the brief text on the map with the extended information provided in the annotations by recognizing the keyword match.

\noindent\textbf{\MakeUppercase{Icon}. }
Icons can also used as visual prompts on the map. These icons represent specific features or data points and are designed to correlate with larger, more detailed icons or illustrations in the explanatory section of the infographic. Icons can be particularly effective in scenarios where a visual representation is more easily recognizable than text.

\noindent\textbf{\MakeUppercase{Color}. }
This technique involves assigning specific colors to map regions and then using the same colors for corresponding labels in the annotations. The color-coding system provides a visual link between the geographic data and the explanatory text, making it simple for viewers to identify which annotation pertains to which map feature. This method can also enhance the overall aesthetic of the infographic by creating a cohesive color scheme.

\subsubsection{Linked}
These labels are connected to their target regions on the map as callouts through a visual link, such as a leader line or an arrow. Linked labels can be further categorized into three types:

\noindent\textbf{\MakeUppercase{Convenient}. }
The labels for this type are strategically placed in open spaces near their corresponding map regions to prioritize ease of reference (see \cref{fig:corpus}.e in Appendix). This positioning ensures that users can quickly identify the data associated with each region without cluttering the map itself. 

\noindent\textbf{\MakeUppercase{Aligned}. }
Labels of this type are arranged around the map's perimeter, creating a neat appearance (see \cref{fig:corpus}.f in Appendix). This approach allows viewers to easily find supplementary information in a specific, consistent place rather than to search around the entire map.  By placing the additional information in fixed areas, such as the top, bottom, or sides of the map where there is more space, this method enhances clarity and accessibility.

\noindent\textbf{\MakeUppercase{Ordered}. }
The deliberate arrangement of labels within the confines of a unified shape can profoundly enhance the narrative or aesthetics of geo-infographic data. Specifically, when depicting forest resources, a collective label configuration can mimic the shape of a tree (see \cref{fig:corpus}.g in Appendix). This technique not only reinforces the thematic message of the data but also organizes the labels in a visually appealing and meaningful way. As a result, these labels may enhance the viewer's understanding and retention of the data's context.

\subsection{Dimension 4: Highlighting Techniques}
A range of highlighting techniques have also been identified to guide viewers' attention towards specific sections of the geo-infographic. 

\setlength\intextsep{0pt}
\setlength\columnsep{0pt}
\begin{wrapfigure}{l}{.12\columnwidth}
    \vspace*{\fill}
    \includegraphics[height=1.8\baselineskip]{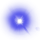}
    \vspace*{\fill}
\end{wrapfigure}
\noindent\textbf{\MakeUppercase{glow}.}
This method uses a luminous point to signify a location, drawing the viewer's attention to a particular area as a focal point. For example, a glowing dot might highlight a city's location on a night-time map.

\setlength\intextsep{0pt}
\setlength\columnsep{0pt}
\begin{wrapfigure}{l}{.12\columnwidth}
    \vspace*{\fill}
    \includegraphics[height=1.8\baselineskip]{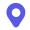}
    \vspace*{\fill}
\end{wrapfigure}
\noindent\textbf{\MakeUppercase{Icon}.}
This method involves placing an icon like a map pin at the exact location of interest, directly drawing the viewer's attention to that point. For instance, this icon can highlight a tourist spot on a city map, making it easy for tourists to navigate.

\setlength\intextsep{0pt}
\setlength\columnsep{0pt}
\begin{wrapfigure}{l}{.12\columnwidth}
    \vspace*{\fill}
    \includegraphics[height=1.8\baselineskip]{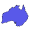}
    \vspace*{\fill}
\end{wrapfigure}
\noindent\textbf{\MakeUppercase{Contrasting Color}.}
This method applies a color that contrasts with the rest of the map to a specific region, highlighting the area of interest. For example, a specific country might be shaded in a bright color to distinguish it from its neighbors on a geopolitical map.

\setlength\intextsep{0pt}
\setlength\columnsep{0pt}
\begin{wrapfigure}{l}{.12\columnwidth}
    \vspace*{\fill}
    \includegraphics[height=1.8\baselineskip]{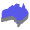}
    \vspace*{\fill}
\end{wrapfigure}
\noindent\textbf{\MakeUppercase{3D Transformation}.}
This method enhances the base map with three-dimensional effects, such as shadows or extrusions, creating a visual illusion of height and prominence.

\setlength\intextsep{0pt}
\setlength\columnsep{0pt}
\begin{wrapfigure}{l}{.12\columnwidth}
    \vspace*{\fill}
    \includegraphics[height=1.8\baselineskip]{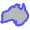}
    \vspace*{\fill}
\end{wrapfigure}
\noindent\textbf{\MakeUppercase{Contour}.}
This method outlines a region on the base map, helping to distinguish areas that are significant.

\setlength\intextsep{0pt}
\setlength\columnsep{0pt}
\begin{wrapfigure}{l}{.12\columnwidth}
    \vspace*{\fill}
    \includegraphics[height=1.8\baselineskip]{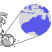}
    \vspace*{\fill}
\end{wrapfigure}
\noindent\textbf{\MakeUppercase{Zoomed Insets}.}
This method involved magnifying a specific map area that is too small to be represented adequately at its actual size. This technique is particularly useful for highlighting areas of great importance that are geographically small, such as an island. The enlarged section can be displayed in an adjacent blank space to avoid obstructing the view of the rest of the map. Alternatively, it can be overlaid directly on its original location, which may partially obscure surrounding areas and reduce the distraction of the viewer's attention.

\begin{figure*}[htb]
 \centering 
 \includegraphics[width=0.85\textwidth]{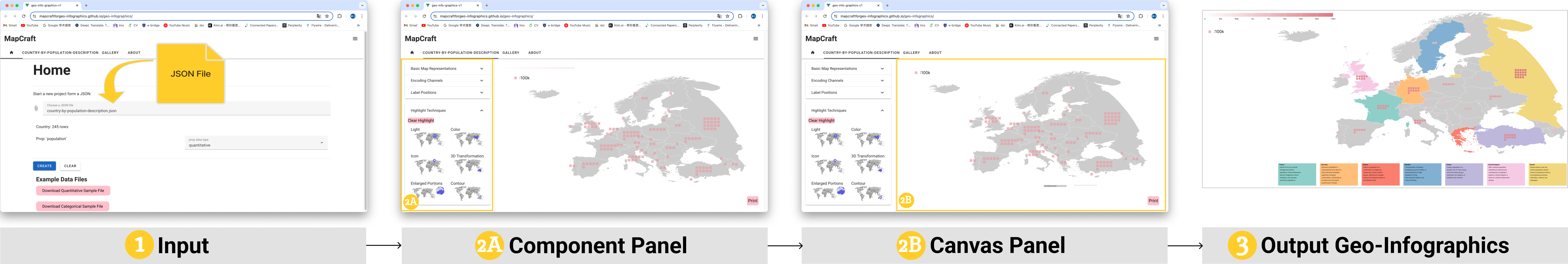}
 \caption{The process of creating geo-infographics using \textit{MapCraft}}
 \label{fig:MapCraft}
\end{figure*}

\section{Geo-infographic Authoring Tool: \textit{MapCraft}}
\subsection{System Design}
Based on the proposed design space, we developed \textit{MapCraft}, a tool designed to streamline the creation of geo-infographics (see \cref{fig:MapCraft}). Users can upload their map-related data, select appropriate encoding channels, choose highlighting techniques, and design and place labels. The tool then allows users to export a comprehensive geo-infographic.

\noindent \textbf{Data Upload Interface.}
Upon initial access, users are presented with the Data Upload Interface (see \cref{fig:MapCraft}.1), where they can import JSON files. These files should be an array of objects, each containing key-value pairs for country name, data to be encoded, and an optional label, \revision{ as demonstrated in \cref{fig:dataFile} in Appendix.}

\noindent \textbf{Authoring Interface.}
The \textit{MapCraft} authoring interface (see \cref{fig:MapCraft}.2) is divided into two components: the Components Panel (see \cref{fig:MapCraft}.2A) and the Canvas Panel (see \cref{fig:MapCraft}.2B).

\begin{itemize}[left=0pt]
    \item \textbf{Components Panel}: Located on the left side, this panel allows users to select visual elements from four key dimensions of geo-infographic design: basic map representations, encoding channels, label design and placement, and highlighting techniques. Users can explore different design choices for each dimension, enhancing their understanding of how various encoding channels and design options affect the effectiveness and aesthetics of the geo-infographic. The ``dual encoding'' feature within the Components Panel allows users to apply multiple encoding channels simultaneously. 
    
    \revision{Based on the type of data to be encoded, the system will have a popup alert to remind users of inapplicable encoding channels and suggest alternative compatible options. This feature helps avoid conflicting encoding channels and assists in creating reasonable geo-infographics, especially for users with limited experience.}
    \item \textbf{Canvas Panel}: Serving as the central workspace, this panel displays real-time updates based on the user's selections from the Components Panel. Users can see immediate visual feedback, enabling them to refine their designs for both effectiveness and visual appeal iteratively.
\end{itemize}

Upon finalizing their design, users can export the geo-infographic as an SVG file using the ``print'' button, capturing the final design from the Canvas Panel.

\noindent \textbf{Gallery and About Interfaces.}
The Gallery Interface (see \cref{fig:gallery} in Appendix) showcases examples created by experienced designers to inspire users, while the About Interface (see \cref{fig:about} in Appendix) provides a brief overview and guidance on using the tool effectively.

\subsection{Implementation.}
\href{https://mapcraftforgeo-infographics.github.io/geo-infographics/}{\textit{MapCraft}} was implemented using D3.js, Vue.js, and Node.js, ensuring a responsive and dynamic user experience.

\section{Evaluation}
To assess the usability and educational value of \textit{MapCraft}, we engaged 12 participants with an interest in visualization but no prior design experience. The participants were asked to create geo-infographics using provided data in three stages: before using the system, while using the system, and after using the system. We collected both quantitative and qualitative data to analyze the educational impact and usability of our tool.

\noindent \textbf{Participants and Apparatus.}
Participants consisted of 12 undergraduate students (6 females and 6 males) aged between 18 and 22, with an average age of 19.5 years (SD = 1.02). All participants were computer science majors from the local university but lacked prior experience in visualization. These participants were specifically chosen as they represent our target user group—individuals who are challenged to create geo-infographics using existing visualization tools.

The tests were conducted in a school computer lab, utilizing AOC CU34G2X monitors, which are 34-inch LCD displays with a resolution of 3440 × 1440 pixels. This setting ensured a consistent and controlled environment for the participants to perform the required tasks.

\begin{figure}[htb]
 \centering
 \includegraphics[width=\columnwidth]{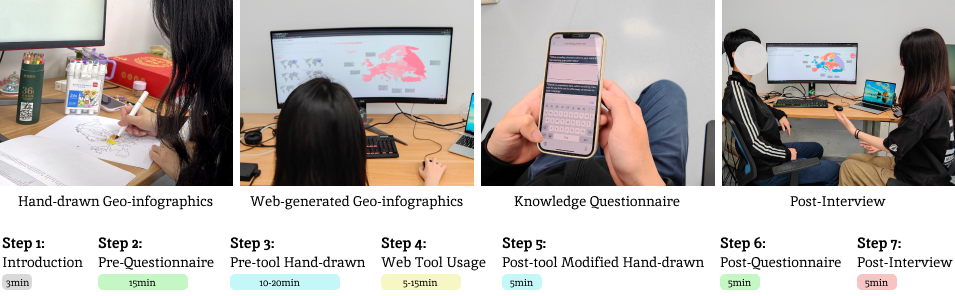}
 \caption{The user study process}
 \label{fig:UserStudyFlowChart}
\end{figure}

\noindent \textbf{Procedure.}
This study was conducted with ethical approval from the university

As \cref{fig:UserStudyFlowChart} shows, the experimental procedure was structured into seven stages, lasting approximately one hour. 
It began with a brief overview of the experimental process and objectives, ensuring participants understood the procedures and aims. This stage also included obtaining consent and informing participants of their rights, including the ability to withdraw at any time, take breaks, and have their personal information protected. 
Next, participants completed a pre-questionnaire to gather background information (see section \ref{sec:background} in Appendix) and assess their initial knowledge of designing geo-infographics. Following this, participants were tasked with drawing two geo-infographics based on provided data: one depicting quantitative population data and the other illustrating categorical country information. Participants then used \textit{MapCraft} to create geo-infographics. After using the tool, participants modified their initial drawings independently. Subsequently, participants completed a post-questionnaire, which included the System Usability Scale (SUS) and the same knowledge assessment as in the pre-questionnaire. Finally, participants were interviewed to discuss their experience using \textit{MapCraft}, provide comments on the design space and usability, and share their intentions for future use.

\begin{figure*}[tb]
 \centering
 \includegraphics[width=0.9\textwidth]{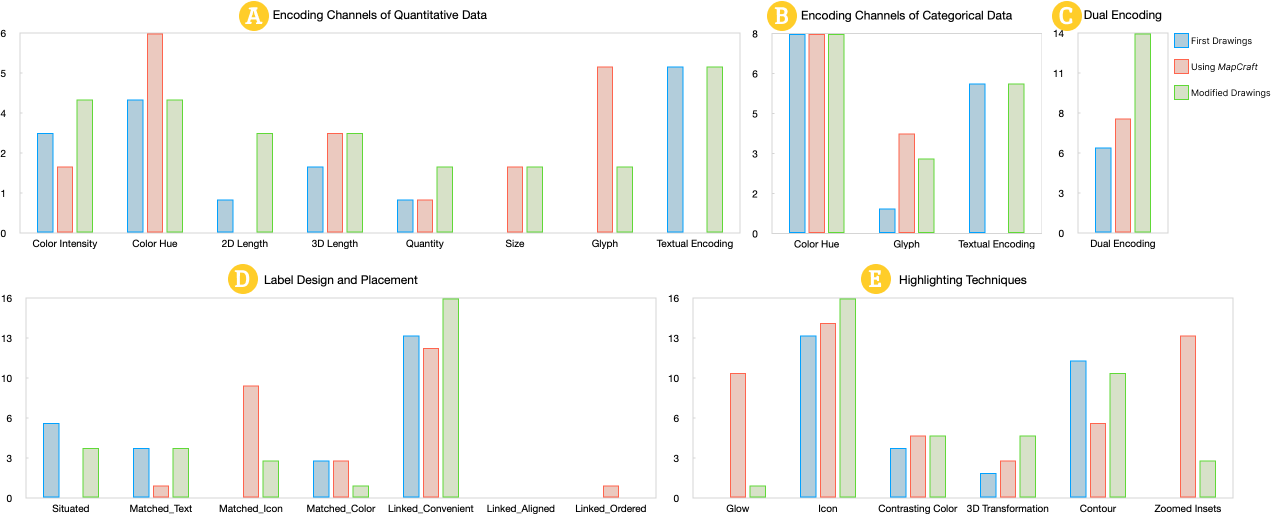}
 \caption{Summary of the frequencies of various design techniques used by participants during the geo-infographic creation process.}
 \label{fig:encodingChannelQuantitative}
\end{figure*}

\noindent \textbf{Data Collection and Analysis.}
We collected various data types to assess the effectiveness and usability of \textit{MapCraft}. Firstly, we gathered the original and modified drafts of geo-infographics created by hand and the graphics designed using \textit{MapCraft} (see \cref{fig:usersGraphic1} and \cref{fig:usersGraphic2} in Appendix), recording the frequency of usage for each method in terms of encoding channels, label design and placement, and highlighting techniques (see \cref{fig:encodingChannelQuantitative}).
Secondly, we measured participants' initial and post-usage knowledge of geo-infographic visualization, including their understanding of encoding channels, label design and placement, highlighting techniques, and dual encoding implementation (see section \ref{sec:knowledge} in Appendix). Thirdly, we recorded the System Usability Scale (SUS) scores to evaluate the usability of \textit{MapCraft}. Lastly, we collected qualitative feedback from the interviews, including participants' comments on the design space and usability, challenges faced while using the tool, and their intentions for future use (see section \ref{sec:interview} in Appendix).

\section{Results and Discussion}

\subsection{Results}

\subsubsection{System Usability Scale (SUS)}
The SUS yielded an overall score of 78.33, indicating a ``Good'' level of usability and placing \textit{MapCraft} within the B grade category.

\subsubsection{Geo-infographics Design}
As depicted in \cref{fig:encodingChannelQuantitative}, we compared the frequency of visual element selection across three stages: hand-drawn geo-infographics prior to \textit{MapCraft} use, geo-infographics created using \textit{MapCraft}, and hand-drawn modifications post-\textit{MapCraft} use. 

\noindent \textbf{First Drawings.}  
In the initial hand-drawn stage, most participants struggled to visualize quantitative and categorical data effectively, preferring single encoding channels. Color was predominantly used for presenting information, and Linked Convenient was the most frequently used label design technique. Highlighting techniques were limited, with users primarily employing Icon and Contour.

\noindent \textbf{Using \textit{MapCraft}.}  
During the use of \textit{MapCraft}, users demonstrated increased diversity in encoding methods, with all provided methods being utilized. Dual encoding became more prevalent and Linked Convenient, and Matched Icon were the most frequently used label design techniques. Highlighting techniques became more varied with Enlarged Portions and Glow, indicating users' exploration of new methods.
\revision{Additionally, all participants took an average of 10 minutes from their first encounter with the system to successfully produce two beautiful geo-infographics, with the longest adaptation time being only 16 minutes.}

\noindent \textbf{Modified Drawings.}  
After using \textit{MapCraft}, participants modified their initial drawings, showing significant improvement in their ability to encode data correctly. Dual encoding usage increased, and users paid greater attention to diverse encoding methods, with Linked Convenient remaining the most popular label design technique. Highlighting techniques became more balanced across different methods, indicating a deeper understanding of possible design choices.

\textbf{Overall}, users demonstrated improved performance on designing geo-infographics after using \textit{MapCraft}. This improvement was reflected in their ability to handle conflicts, utilize dual encoding, and diversify visualization methods.

\subsubsection{Knowledge Assessment.}
We analyzed the knowledge level from pre- and post-questionnaires, focusing on the methods users employ to design geo-infographics.

\noindent \textbf{Before Using \textit{MapCraft}}  
Five participants struggled with encoding channels in the pre-questionnaire, and none could apply dual encoding methods correctly. Method selection in label design, placement, and highlighting techniques were very limited. Most users could present labels and add highlights, but often, their methods were inappropriate, such as using color for both encoding and highlighting simultaneously. Overall, users relied on basic methods, frequently using color for highlighting and simply filling the map with color rather than exploring various applications.

\noindent \textbf{After Using \textit{MapCraft}}  
After using \textit{MapCraft}, all participants could select one appropriate and correct method in these dimensions, with only one participant still confused about dual encoding. Participants began demonstrating creativity in highlighting techniques and combining various methods.

The comparison of knowledge before and after using \textit{MapCraft} indicates that users showed enhanced competence in selecting and utilizing methods. Their design thoughts became more mature and feasible. The findings suggested users enriched their knowledge of geographic visualization methods after using our system. Participants were eager to explore and apply newly acquired, novel, and visually appealing techniques, leading to increased diversity in design method choices in the post-questionnaire responses.

\subsubsection{Interview.}
Participants (P) endorsed the four-dimensional framework of our proposed design space for geo-infographic creation, describing it as logical, clear, and intuitive. For instance, P5 remarked that ``dividing the creation of a geo-infographic into these four parts is logically sound,'' and P2 noted, ``You start with a map and then add elements one by one.'' This feedback highlights the framework's accessibility and ease of use, even for individuals lacking prior experience in data visualization.
Several participants emphasized \textit{MapCraft}'s ability to integrate various visual elements seamlessly. P3 mentioned the tool's capacity to ``combine a variety of visual elements, to arrange and piece them together,'' while P9 appreciated how it allows users to ``draw a map that can convey different pieces of information, integrating them seamlessly.''
Participants agreed that \textit{MapCraft} is highly user-friendly, convenient, and efficient. Participants such as P2 and P3 found it ``much more convenient than hand-drawing or searching for other tools,'' and P5 praised its ability to ``quickly represent the information of countries and create a visually appealing map.'' P12 and P13 highlighted its convenience and practical functionality, with P14 describing it as ``quite practical.''
\revision{However, P5 and P6 noted the need for extra instruction when prompts appear to avoid illegal operations. Additionally, some users require more exploration time and effort before they can appropriately apply the features. Regarding the dual encoding function, most users were attracted by its sophistication and were surprised by this novel visualization method. However, this new approach also introduced some learning costs. As P8 mentioned, the feature requires only ``limited understanding'' to get started, and as P14 stated, ``,It's a little challenging but quite fun.''}

Overall, the positive feedback suggests that the four-dimensional design space effectively guides users through the geo-infographic creation process, making it accessible and straightforward. \textit{MapCraft} enables even novice users to produce aesthetically pleasing and informative geo-infographics efficiently. 
Participants expressed strong intentions to utilize \textit{MapCraft} for future geo-infographic creation, underscoring its practical value and usability.

\subsection{Discussion}
\subsubsection{Limitation}

Following Tullis's research \cite{SUSreliable}, the SUS is a widely recognized method for assessing usability, providing a reliable score even with smaller sample sizes, such as 12 to 14 participants. Our participant pool of 12 lends credibility to our findings. However, 12 samples may be sufficient for a comprehensive evaluation in other aspects, such as the knowledge test.
While we maintained gender balance in our participant group of 6 males and 6 females, the age range was narrowly defined from 18 to 30, excluding participants from other age groups. Additionally, all participants were university-educated, either as undergraduates or postgraduates.
This focus may have excluded perspectives from a broader, potentially less educated user base. However, considering our tool is for creating geo-infographics, it is arguable whether lower education levels are within our target demographic.

Furthermore, we reflect on our data analysis methodology, which focused on the diversity of participants' designs and creative ideas. We did not rigorously measure their accuracy in understanding various encoding channels. As an educational tool, future iterations may benefit from more serious assessments of educational outcomes. This could involve pre- and post-tests designed to evaluate the effectiveness of different encoding channels, ensuring a more thorough evaluation of the tool's educational impact.

\subsubsection{Geo-Infographics vs Interactive Map Visualization}

While utilizing geographic information, interactive map applications and geo-infographics serve different purposes and have distinct design requirements. We explored whether our design space for geo-infographics could be applied to interactive map visualizations.
Interactive map applications like Google Maps\footnote{\url{https://maps.google.com/}} are dynamic platforms designed for real-time interaction and navigation, prioritizing accuracy over aesthetics. Conversely, geo-infographics are static visual representations focused on clarity, visual appeal, and storytelling.
Both formats share similarities in basic maps, encoding channels, and highlighting techniques. In geo-infographics, encoding channels and basic maps are chosen for aesthetic appeal and accuracy, whereas in interactive maps, accuracy tends to be prioritized. As a result, while the same basic maps, encoding channels and highlighting techniques may be used in both formats, their selection and frequency can differ based on these priorities.
Another significant difference lies in label design and placement. Geo-infographics often use various label types, including linked labels, to ensure all information is visible within the static layout. In contrast, interactive maps typically use interactive labels that appear upon user action, such as clicks or hovers, reducing the need for complex label arrangements and allowing for a cleaner interface.

\subsubsection{Authoring and Education}

 In evaluating \textit{MapCraft}, we assessed the changes in participants' knowledge about geo-infographics and their ability to generate them. From positive results, we can define our tool as both an educational and an authoring tool, breaking traditional boundaries between these two categories.

\noindent 1) \textit{Comparison with Purely Educational and Purely Authoring Tools.} 
\noindent\textbf{Purely Educational Tools}: These tools are typically designed to convey information and educate users about specific topics. They often use demonstrations, tutorials, or static examples to illustrate concepts. In a geo-infographics context, a purely educational tool might display various dimensions of geo-infographic design — such as encoding channels, map types, and labeling techniques — without providing an interactive or hands-on component. The primary limitation of these tools is the lack of user engagement. Users may understand theoretical aspects but lack the opportunity to apply this knowledge in a practical setting, which can hinder deeper learning and retention.

\noindent \textbf{Purely Authoring Tools}: These tools are designed to automate the creation process, often providing users with a set of inputs and generating visualizations automatically. They prioritize efficiency and ease of use. For geo-infographics, a purely authoring tool might allow users to input data and automatically generate a geo-infographic without requiring any design decisions from the user. The main drawback of these tools is their limited educational value. While they can produce visualizations quickly, users do not gain an understanding of the underlying design principles or improve their visualization skills through the process. For example, a powerful map visualization tool, Flourish\footnote{\url{https://app.flourish.studio/projects}}, is often employed for map visualization and storytelling. It provides powerful creation tools but does not allow users to learn how to design geo-infographics. Inexperienced users may struggle to gain visualization knowledge and improve their design skills while using this tool.

\noindent 2) \textit{Balancing Automation and User Involvement.}
The choice between purely automated tools and those requiring significant user involvement often depends on the target users and their needs:
\begin{itemize}[left=0pt]
    \item \textbf{Automated Tools}: These are ideal for users who need quick results without deep engagement in the design process. They are suitable for scenarios where efficiency is paramount, such as in business environments where time constraints are critical, or for users with limited visualization expertise who need to generate professional-looking results with minimal effort. In the system developed like DataShot \cite{Text-to-Viz}, infographics are generated by the system with pre-designed styles, sparing users the need to spend a significant amount of time learning or possessing visualization design experience. Users only need to make simple edits to obtain a professional infographic.
    \item \textbf{User-Involved Tools}: These tools benefit users who seek to understand and learn from the visualization process. They are suited for educational settings, training programs, and scenarios where the quality and customization of the final product are more important than the creation speed. Users in these contexts benefit from the hands-on experience and the opportunity to apply theoretical knowledge in practice. Experienced users can create highly personalized and professional geo-infographics, while inexperienced users can also gain design experience with guidance from official documentation.
\end{itemize}

However, achieving the right balance between automation and user involvement is crucial. More automated tools might be preferable for professional environments where quick turnaround and consistency are essential. In contrast, educational tools benefit from greater user involvement, which fosters a deeper understanding of the subject matter. 

\section{Conclusion}
This research started with an extensive analysis of 118 geo-infographics found online and design ideas from eight visualization researchers. From these efforts, we developed a comprehensive design space for geo-infographic design. Based on the design space, we created \textit{MapCraft}, a web-based authoring tool designed to simplify geo-infographics creation.
Our user study with 12 potential users demonstrated that \textit{MapCraft} effectively assists users in creating high-quality geo-infographics while enhancing their understanding of diverse geo-infographic design choices. The tool was found to be user-friendly, straightforward, and valuable for users.
Additionally, future advancements could focus on increasing automation and intelligence within the tool, thereby reducing the need for extensive user education and enabling faster infographic creation.

\acknowledgments{
This work was supported in part by a grant from RDF-22-01-092.}
\bibliographystyle{abbrv-doi-narrow}
\bibliography{template}

\begin{thebibliography}{10}
\renewcommand*{\sfdefault}{PTSansNarrow-TLF}

\bibitem{10.1145/3377325.3377517}
O.~Barral, S.~Lall\'{e}, and C.~Conati.
\newblock Understanding the effectiveness of adaptive guidance for narrative visualization: a gaze-based analysis.
\newblock In {\em Proceedings of the 25th International Conference on Intelligent User Interfaces}, IUI '20, p. 1–9. Association for Computing Machinery, New York, NY, USA, 2020. doi: \textsf{%
10\hspace{.1pt}\discretionary{.}{%
}{.}\hspace{.4pt}1145\discretionary{/}{%
}{/}3377325\hspace{.1pt}\discretionary{.}{%
}{.}\hspace{.4pt}3377517}


\bibitem{BRAIN733}
H.~Bicen and M.~Beheshti.
\newblock The psychological impact of infographics in education.
\newblock {\em BRAIN. Broad Research in Artificial Intelligence and Neuroscience}, 8(4):99--108, 2017.

\bibitem{Glyph-based}
R.~Borgo, J.~Kehrer, D.~H.~S. Chung, E.~Maguire, R.~S. Laramee, H.~Hauser, M.~Ward, and M.~Chen.
\newblock {Glyph-based Visualization: Foundations, Design Guidelines, Techniques and Applications}.
\newblock In M.~Sbert and L.~Szirmay-Kalos, eds., {\em Eurographics 2013 - State of the Art Reports}. The Eurographics Association, 2013. doi: \textsf{%
10\hspace{.1pt}\discretionary{.}{%
}{.}\hspace{.4pt}2312\discretionary{/}{%
}{/}conf\discretionary{/}{%
}{/}EG2013\discretionary{/}{%
}{/}stars\discretionary{/}{%
}{/}039\discretionary{%
}{-}{-}063}


\bibitem{ScatterBlogs2}
H.~Bosch, D.~Thom, F.~Heimerl, E.~Püttmann, S.~Koch, R.~Krüger, M.~Wörner, and T.~Ertl.
\newblock Scatterblogs2: Real-time monitoring of microblog messages through user-guided filtering.
\newblock {\em IEEE Transactions on Visualization and Computer Graphics}, 19(12):2022--2031, 2013. doi: \textsf{%
10\hspace{.1pt}\discretionary{.}{%
}{.}\hspace{.4pt}1109\discretionary{/}{%
}{/}TVCG\hspace{.1pt}\discretionary{.}{%
}{.}\hspace{.4pt}2013\hspace{.1pt}\discretionary{.}{%
}{.}\hspace{.4pt}186}


\bibitem{TimelinesRevisited}
M.~Brehmer, B.~Lee, B.~Bach, N.~H. Riche, and T.~Munzner.
\newblock Timelines revisited: A design space and considerations for expressive storytelling.
\newblock {\em IEEE Transactions on Visualization and Computer Graphics}, 23(9):2151--2164, 2017. doi: \textsf{%
10\hspace{.1pt}\discretionary{.}{%
}{.}\hspace{.4pt}1109\discretionary{/}{%
}{/}TVCG\hspace{.1pt}\discretionary{.}{%
}{.}\hspace{.4pt}2016\hspace{.1pt}\discretionary{.}{%
}{.}\hspace{.4pt}2614803}


\bibitem{VAUD}
W.~Chen, Z.~Huang, F.~Wu, M.~Zhu, H.~Guan, and R.~Maciejewski.
\newblock Vaud: A visual analysis approach for exploring spatio-temporal urban data.
\newblock {\em IEEE Transactions on Visualization and Computer Graphics}, 24(9):2636--2648, 2018. doi: \textsf{%
10\hspace{.1pt}\discretionary{.}{%
}{.}\hspace{.4pt}1109\discretionary{/}{%
}{/}TVCG\hspace{.1pt}\discretionary{.}{%
}{.}\hspace{.4pt}2017\hspace{.1pt}\discretionary{.}{%
}{.}\hspace{.4pt}2758362}


\bibitem{Text-to-Viz}
W.~Cui, X.~Zhang, Y.~Wang, H.~Huang, B.~Chen, L.~Fang, H.~Zhang, J.-G. Lou, and D.~Zhang.
\newblock Text-to-viz: Automatic generation of infographics from proportion-related natural language statements.
\newblock {\em IEEE Transactions on Visualization and Computer Graphics}, 26(1):906--916, 2020. doi: \textsf{%
10\hspace{.1pt}\discretionary{.}{%
}{.}\hspace{.4pt}1109\discretionary{/}{%
}{/}TVCG\hspace{.1pt}\discretionary{.}{%
}{.}\hspace{.4pt}2019\hspace{.1pt}\discretionary{.}{%
}{.}\hspace{.4pt}2934785}


\bibitem{doi:10.1177/09520767221140954}
S.~Dailey, B.~Gilmore, and N.~Rangarajan.
\newblock The visualization of public information: Describing the use of narrative infographics by u.s. municipal governments.
\newblock {\em Public Policy and Administration}, 0(0):09520767221140954, 0. doi: \textsf{%
10\hspace{.1pt}\discretionary{.}{%
}{.}\hspace{.4pt}1177\discretionary{/}{%
}{/}09520767221140954}


\bibitem{gastner2004diffusion}
M.~T. Gastner and M.~E.~J. Newman.
\newblock Diffusion-based method for producing density-equalizing maps.
\newblock {\em Proceedings of the National Academy of Sciences}, 101(20):7499--7504, 2004. doi: \textsf{%
10\hspace{.1pt}\discretionary{.}{%
}{.}\hspace{.4pt}1073\discretionary{/}{%
}{/}pnas\hspace{.1pt}\discretionary{.}{%
}{.}\hspace{.4pt}0400280101}


\bibitem{gebre2016developing}
E.~H. Gebre and J.~L. Polman.
\newblock Developing young adults' representational competence through infographic-based science news reporting.
\newblock {\em International Journal of Science Education}, 38(18):2667--2687, 2016. doi: \textsf{%
10\hspace{.1pt}\discretionary{.}{%
}{.}\hspace{.4pt}1080\discretionary{/}{%
}{/}09500693\hspace{.1pt}\discretionary{.}{%
}{.}\hspace{.4pt}2016\hspace{.1pt}\discretionary{.}{%
}{.}\hspace{.4pt}1258129}


\bibitem{Graser2016}
A.~Graser.
\newblock {\em Learning QGIS}.
\newblock Packt Publishing Ltd, 2016.

\bibitem{harrison2015infographic}
L.~Harrison, K.~Reinecke, and R.~Chang.
\newblock Infographic aesthetics: Designing for the first impression.
\newblock In {\em Proceedings of the 33rd Annual ACM Conference on Human Factors in Computing Systems}, CHI '15, p. 1187–1190. Association for Computing Machinery, New York, NY, USA, 2015. doi: \textsf{%
10\hspace{.1pt}\discretionary{.}{%
}{.}\hspace{.4pt}1145\discretionary{/}{%
}{/}2702123\hspace{.1pt}\discretionary{.}{%
}{.}\hspace{.4pt}2702545}


\bibitem{he_visualize_2011}
M.~He, X.~Tang, and Y.~Huang.
\newblock To visualize spatial data using thematic maps combined with infographics.
\newblock In {\em 2011 19th International Conference on Geoinformatics}, pp. 1--5. IEEE, Shanghai, China, 2011. doi: \textsf{%
10\hspace{.1pt}\discretionary{.}{%
}{.}\hspace{.4pt}1109\discretionary{/}{%
}{/}GeoInformatics\hspace{.1pt}\discretionary{.}{%
}{.}\hspace{.4pt}2011\hspace{.1pt}\discretionary{.}{%
}{.}\hspace{.4pt}5980880}


\bibitem{https://doi.org/10.1111/cgf.14031}
M.~Hogräfer, M.~Heitzler, and H.-J. Schulz.
\newblock The state of the art in map-like visualization.
\newblock {\em Computer Graphics Forum}, 39(3):647--674, 2020. doi: \textsf{%
10\hspace{.1pt}\discretionary{.}{%
}{.}\hspace{.4pt}1111\discretionary{/}{%
}{/}cgf\hspace{.1pt}\discretionary{.}{%
}{.}\hspace{.4pt}14031}


\bibitem{Lei2023}
F.~Lei, Y.~Ma, A.~S. Fotheringham, et~al.
\newblock Geoexplainer: A visual analytics framework for spatial modeling contextualization and report generation.
\newblock {\em IEEE Transactions on Visualization and Computer Graphics}, 2023.

\bibitem{lerman2021evaluation}
S.~Lerman~Ginzburg, P.~Botana~Martinez, E.~Reisner, S.~Chappell, D.~Brugge, and S.~Kurtz-Rossi.
\newblock An evaluation of an environmental health infographic in community settings.
\newblock {\em INQUIRY: The Journal of Health Care Organization, Provision, and Financing}, 58:00469580211059290, 2021. doi: \textsf{%
10\hspace{.1pt}\discretionary{.}{%
}{.}\hspace{.4pt}1177\discretionary{/}{%
}{/}00469580211059290}


\bibitem{GeoCamera}
W.~Li, Z.~Wang, Y.~Wang, D.~Weng, L.~Xie, S.~Chen, H.~Zhang, and H.~Qu.
\newblock Geocamera: Telling stories in geographic visualizations with camera movements.
\newblock In {\em Proceedings of the 2023 CHI Conference on Human Factors in Computing Systems}, CHI '23. Association for Computing Machinery, New York, NY, USA, 2023. doi: \textsf{%
10\hspace{.1pt}\discretionary{.}{%
}{.}\hspace{.4pt}1145\discretionary{/}{%
}{/}3544548\hspace{.1pt}\discretionary{.}{%
}{.}\hspace{.4pt}3581470}


\bibitem{VIF}
M.~Lu, C.~Wang, J.~Lanir, N.~Zhao, H.~Pfister, D.~Cohen-Or, and H.~Huang.
\newblock Exploring visual information flows in infographics.
\newblock In {\em Proceedings of the 2020 CHI Conference on Human Factors in Computing Systems}, CHI '20, p. 1–12. Association for Computing Machinery, New York, NY, USA, 2020. doi: \textsf{%
10\hspace{.1pt}\discretionary{.}{%
}{.}\hspace{.4pt}1145\discretionary{/}{%
}{/}3313831\hspace{.1pt}\discretionary{.}{%
}{.}\hspace{.4pt}3376263}


\bibitem{mansour2022use}
E.~Mansour.
\newblock Use of infographics as a technology-based information dissemination tool: the perspective of egyptian public university libraries library staff.
\newblock {\em Library Hi Tech}, 40(6):1819--1842, 2022.

\bibitem{MARTIN201948}
L.~J. Martin, A.~Turnquist, B.~Groot, S.~Y. Huang, E.~Kok, B.~Thoma, and J.~J. {van Merriënboer}.
\newblock Exploring the role of infographics for summarizing medical literature.
\newblock {\em Health Professions Education}, 5(1):48--57, 2019. doi: \textsf{%
10\hspace{.1pt}\discretionary{.}{%
}{.}\hspace{.4pt}1016\discretionary{/}{%
}{/}j\hspace{.1pt}\discretionary{.}{%
}{.}\hspace{.4pt}hpe\hspace{.1pt}\discretionary{.}{%
}{.}\hspace{.4pt}2018\hspace{.1pt}\discretionary{.}{%
}{.}\hspace{.4pt}03\hspace{.1pt}\discretionary{.}{%
}{.}\hspace{.4pt}005}


\bibitem{munzner2014visualization}
T.~Munzner.
\newblock Visualization analysis and design: keynote address.
\newblock {\em J. Comput. Sci. Coll.}, 32(1):106–107, oct 2016.

\bibitem{Naparin2017}
H.~Naparin and A.~B. Saad.
\newblock Infographics in education: Review on infographics design.
\newblock {\em The International Journal of Multimedia \& Its Applications (IJMA)}, 9(4):5, 2017.

\bibitem{Ormsby2004}
T.~Ormsby.
\newblock {\em Getting to Know ArcGIS Desktop: Basics of ArcView, ArcEditor, and ArcInfo}.
\newblock ESRI, Inc., 2004.

\bibitem{peterson2020gis}
G.~Peterson.
\newblock {\em GIS Cartography: A Guide to Effective Map Design}.
\newblock 10 2020. doi: \textsf{%
10\hspace{.1pt}\discretionary{.}{%
}{.}\hspace{.4pt}1201\discretionary{/}{%
}{/}9781003046325}


\bibitem{InformationGraphics}
S.~Rendgen.
\newblock {\em Information graphics}.
\newblock Taschen, K{\"o}ln, 2012.
\newblock Beilage unter dem Titel: Holmes, Nigel: Nigel Holmes' map of infographia: an idiosyncratic taxonomy.

\bibitem{roth2021cartographic}
R.~E. Roth.
\newblock Cartographic design as visual storytelling: synthesis and review of map-based narratives, genres, and tropes.
\newblock {\em The Cartographic Journal}, 58(1):83--114, 2021. doi: \textsf{%
10\hspace{.1pt}\discretionary{.}{%
}{.}\hspace{.4pt}1080\discretionary{/}{%
}{/}00087041\hspace{.1pt}\discretionary{.}{%
}{.}\hspace{.4pt}2019\hspace{.1pt}\discretionary{.}{%
}{.}\hspace{.4pt}1633103}


\bibitem{Song_Roth_Houtman_Prestby_Iverson_Gao_2022}
Z.~Song, R.~E. Roth, L.~Houtman, T.~Prestby, A.~Iverson, and S.~Gao.
\newblock Visual storytelling with maps: An empirical study on story map themes and narrative elements, visual storytelling genres and tropes, and individual audience differences.
\newblock {\em Cartographic Perspectives}, (100):10–44, Aug. 2022. doi: \textsf{%
10\hspace{.1pt}\discretionary{.}{%
}{.}\hspace{.4pt}14714\discretionary{/}{%
}{/}CP100\hspace{.1pt}\discretionary{.}{%
}{.}\hspace{.4pt}1759}


\bibitem{PB-VRVis-2023-023}
R.~Splechtna, T.~Hulka, D.~Sardana, N.~D. Chandrashekar, D.~Gra{\v{c}}anin, and K.~Matkovi{\'c}.
\newblock Interactive exploration of complex heterogeneous data: A use case on understanding city economics.
\newblock In n.n., ed., {\em VISIGRAPP 2023 - 18th International Joint Conference on Computer Vision, Imaging and Computer Graphics Theory and Applications}, 2023. doi: \textsf{%
10\hspace{.1pt}\discretionary{.}{%
}{.}\hspace{.4pt}5220\discretionary{/}{%
}{/}0011787500003417}


\bibitem{tobler2004thirty}
W.~Tobler.
\newblock Thirty five years of computer cartograms.
\newblock {\em ANNALS of the Association of American Geographers}, 94(1):58--73, 2004. doi: \textsf{%
10\hspace{.1pt}\discretionary{.}{%
}{.}\hspace{.4pt}1111\discretionary{/}{%
}{/}j\hspace{.1pt}\discretionary{.}{%
}{.}\hspace{.4pt}1467\discretionary{%
}{-}{-}8306\hspace{.1pt}\discretionary{.}{%
}{.}\hspace{.4pt}2004\hspace{.1pt}\discretionary{.}{%
}{.}\hspace{.4pt}09401004\hspace{.1pt}\discretionary{.}{%
}{.}\hspace{.4pt}x}


\bibitem{SUSreliable}
T.~Tullis and J.~Stetson.
\newblock A comparison of questionnaires for assessing website usability.
\newblock 06 2006.

\bibitem{ManyEyes}
F.~B. Viegas, M.~Wattenberg, F.~van Ham, J.~Kriss, and M.~McKeon.
\newblock Manyeyes: a site for visualization at internet scale.
\newblock {\em IEEE Transactions on Visualization and Computer Graphics}, 13(6):1121--1128, 2007. doi: \textsf{%
10\hspace{.1pt}\discretionary{.}{%
}{.}\hspace{.4pt}1109\discretionary{/}{%
}{/}TVCG\hspace{.1pt}\discretionary{.}{%
}{.}\hspace{.4pt}2007\hspace{.1pt}\discretionary{.}{%
}{.}\hspace{.4pt}70577}


\bibitem{ijerph19159634}
Q.~Wang, K.~Yang, L.~Li, and Y.~Zhu.
\newblock Assessing the terrain gradient effect of landscape ecological risk in the dianchi lake basin of china using geo-information tupu method.
\newblock {\em International Journal of Environmental Research and Public Health}, 19(15), 2022. doi: \textsf{%
10\hspace{.1pt}\discretionary{.}{%
}{.}\hspace{.4pt}3390\discretionary{/}{%
}{/}ijerph19159634}


\bibitem{DataShot}
Y.~Wang, Z.~Sun, H.~Zhang, W.~Cui, K.~Xu, X.~Ma, and D.~Zhang.
\newblock Datashot: Automatic generation of fact sheets from tabular data.
\newblock {\em IEEE Transactions on Visualization and Computer Graphics}, 26(1):895--905, 2020. doi: \textsf{%
10\hspace{.1pt}\discretionary{.}{%
}{.}\hspace{.4pt}1109\discretionary{/}{%
}{/}TVCG\hspace{.1pt}\discretionary{.}{%
}{.}\hspace{.4pt}2019\hspace{.1pt}\discretionary{.}{%
}{.}\hspace{.4pt}2934398}


\bibitem{GraphicsGrammar}
H.~Wickham.
\newblock A layered grammar of graphics.
\newblock {\em Journal of Computational and Graphical Statistics}, 19(1):3--28, 2010. doi: \textsf{%
10\hspace{.1pt}\discretionary{.}{%
}{.}\hspace{.4pt}1198\discretionary{/}{%
}{/}jcgs\hspace{.1pt}\discretionary{.}{%
}{.}\hspace{.4pt}2009\hspace{.1pt}\discretionary{.}{%
}{.}\hspace{.4pt}07098}


\bibitem{Wieczorek2009}
W.~F. Wieczorek and A.~M. Delmerico.
\newblock Geographic information systems.
\newblock {\em Wiley Interdisciplinary Reviews: Computational Statistics}, 1(2):167--186, 2009. doi: \textsf{%
10\hspace{.1pt}\discretionary{.}{%
}{.}\hspace{.4pt}1002\discretionary{/}{%
}{/}wics\hspace{.1pt}\discretionary{.}{%
}{.}\hspace{.4pt}35}


\bibitem{land12061242}
L.~Wu, Y.~Yang, H.~Yang, B.~Xie, and W.~Luo.
\newblock A comparative study on land use/land cover change and topographic gradient effect between mountains and flatlands of southwest china.
\newblock {\em Land}, 12(6), 2023. doi: \textsf{%
10\hspace{.1pt}\discretionary{.}{%
}{.}\hspace{.4pt}3390\discretionary{/}{%
}{/}land12061242}


\bibitem{young2014infographics}
A.~Young and M.~Hinesly.
\newblock Infographics as a business communication tool: An empirical investigation of user preference, comprehension \& efficiency.
\newblock {\em Comprehension \& Efficiency (January 12, 2014)}, 2014. doi: \textsf{%
10\hspace{.1pt}\discretionary{.}{%
}{.}\hspace{.4pt}2139\discretionary{/}{%
}{/}ssrn\hspace{.1pt}\discretionary{.}{%
}{.}\hspace{.4pt}2548559}


\bibitem{Zeng2017}
W.~Zeng et~al.
\newblock Visualizing the relationship between human mobility and points of interest.
\newblock {\em IEEE Transactions on Intelligent Transportation Systems}, 18(8):2271--2284, 2017.

\bibitem{FlowLine}
Y.~Zhao, X.~Tang, F.~Sun, Y.~Zhang, and Y.~Wu.
\newblock Optimization strategies of flow line design in geo-infographics.
\newblock In {\em 2018 26th International Conference on Geoinformatics}, pp. 1--6, 2018. doi: \textsf{%
10\hspace{.1pt}\discretionary{.}{%
}{.}\hspace{.4pt}1109\discretionary{/}{%
}{/}GEOINFORMATICS\hspace{.1pt}\discretionary{.}{%
}{.}\hspace{.4pt}2018\hspace{.1pt}\discretionary{.}{%
}{.}\hspace{.4pt}8557186}


\bibitem{ZHOU2020244}
Z.~Zhou, X.~Zhang, Z.~Guo, and Y.~Liu.
\newblock Visual abstraction and exploration of large-scale geographical social media data.
\newblock {\em Neurocomputing}, 376:244--255, 2020. doi: \textsf{%
10\hspace{.1pt}\discretionary{.}{%
}{.}\hspace{.4pt}1016\discretionary{/}{%
}{/}j\hspace{.1pt}\discretionary{.}{%
}{.}\hspace{.4pt}neucom\hspace{.1pt}\discretionary{.}{%
}{.}\hspace{.4pt}2019\hspace{.1pt}\discretionary{.}{%
}{.}\hspace{.4pt}10\hspace{.1pt}\discretionary{.}{%
}{.}\hspace{.4pt}072}


\bibitem{ZHU202024}
S.~Zhu, G.~Sun, Q.~Jiang, M.~Zha, and R.~Liang.
\newblock A survey on automatic infographics and visualization recommendations.
\newblock {\em Visual Informatics}, 4(3):24--40, 2020. doi: \textsf{%
10\hspace{.1pt}\discretionary{.}{%
}{.}\hspace{.4pt}1016\discretionary{/}{%
}{/}j\hspace{.1pt}\discretionary{.}{%
}{.}\hspace{.4pt}visinf\hspace{.1pt}\discretionary{.}{%
}{.}\hspace{.4pt}2020\hspace{.1pt}\discretionary{.}{%
}{.}\hspace{.4pt}07\hspace{.1pt}\discretionary{.}{%
}{.}\hspace{.4pt}002}


\bibitem{chen2019towards}
C.~Zhu-Tian, Y.~Wang, Q.~Wang, Y.~Wang, and H.~Qu.
\newblock Towards automated infographic design: Deep learning-based auto-extraction of extensible timeline.
\newblock {\em IEEE Transactions on Visualization and Computer Graphics}, 26(1):917--926, 2020. doi: \textsf{%
10\hspace{.1pt}\discretionary{.}{%
}{.}\hspace{.4pt}1109\discretionary{/}{%
}{/}TVCG\hspace{.1pt}\discretionary{.}{%
}{.}\hspace{.4pt}2019\hspace{.1pt}\discretionary{.}{%
}{.}\hspace{.4pt}2934810}


\end{thebibliography}

\clearpage

\appendix
\label{sec:supplemental_materials}
\onecolumn

\begin{center}
\Large{APPENDIX}
\end{center}

\section{Corpus}
\begin{figure}[htb]
    \centering
    \includegraphics[width=\columnwidth]{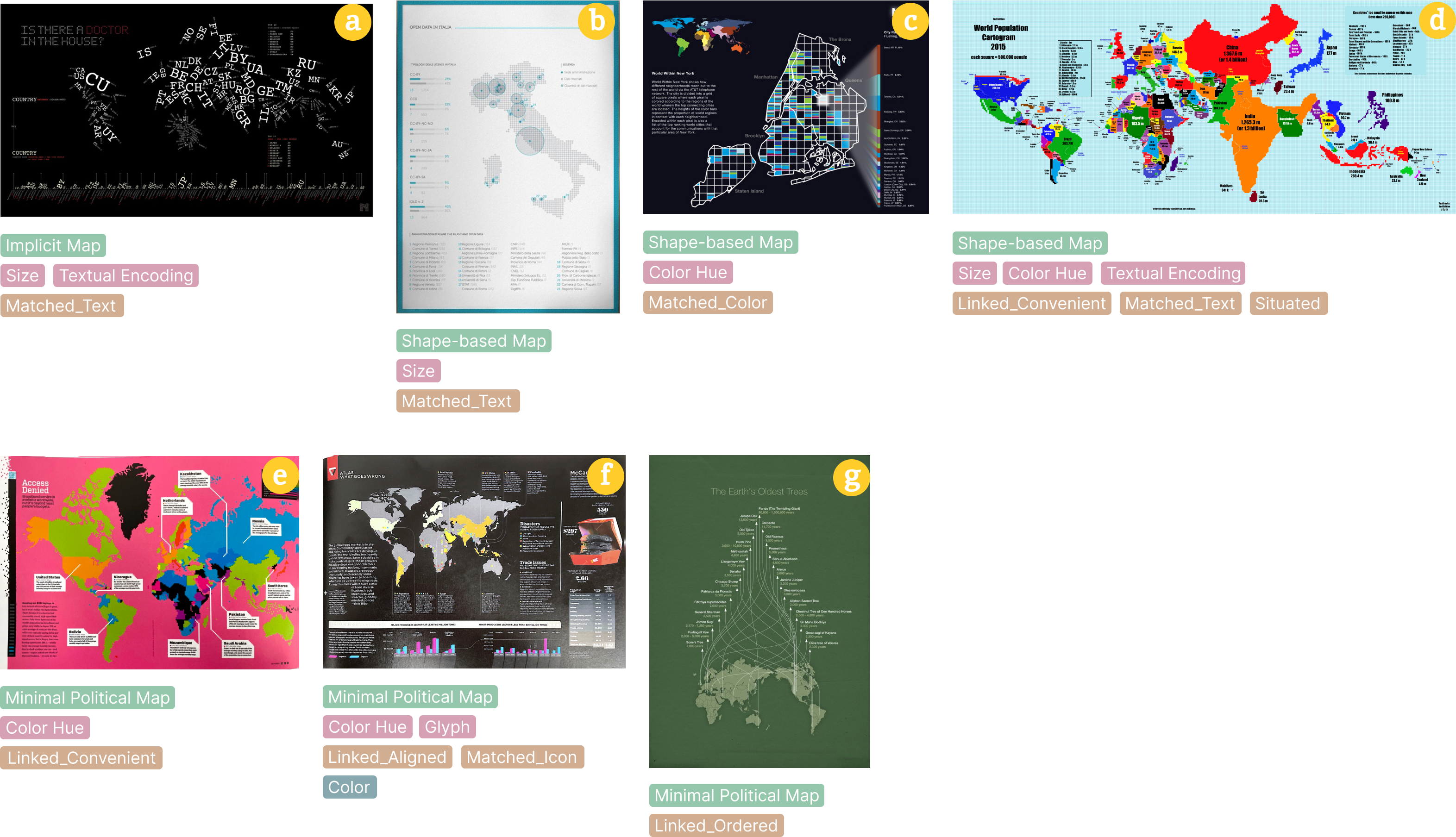}
    \caption{Example geo-infographics in Corpus}
    \label{fig:corpus}
\end{figure}

\cref{fig:corpus} gives examples of different map representations.
\label{sec:corpus}

\section{Complementary user interfaces}

\subsection{Example of Uploaded Data File}
\begin{figure}[htb]
    \centering
    \includegraphics[width=\columnwidth]{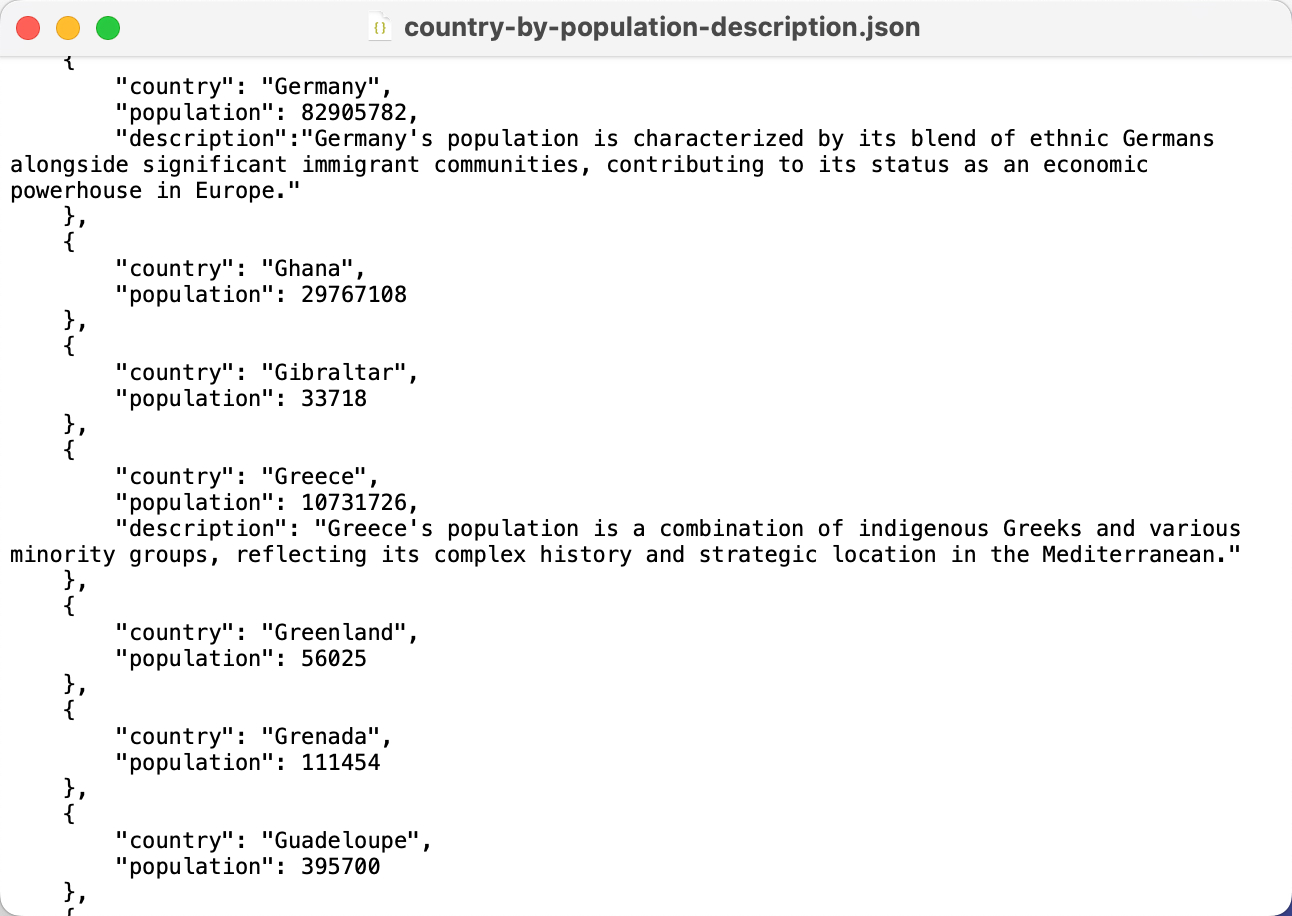}
    \caption{Example of Supported JSON File}
    \label{fig:dataFile}
\end{figure}

\cref{fig:dataFile} shows a screenshot of a part of the data file that is applicable to the system.
\label{sec:dataFile}

\subsection{\MakeUppercase{gallery} Interface}
\begin{figure}[htb]
    \centering
    \includegraphics[width=\columnwidth]{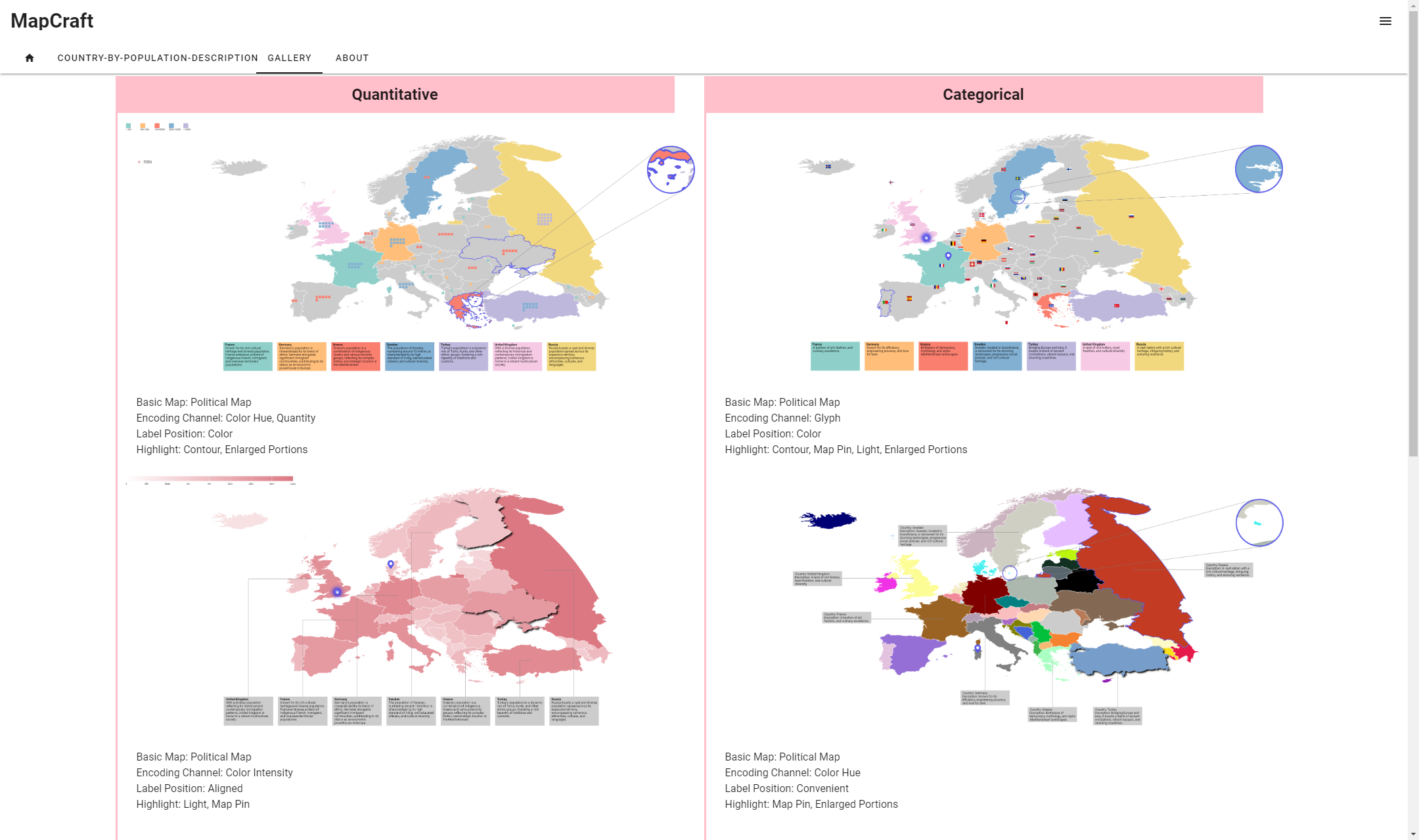}
    \caption{\MakeUppercase{gallery} Interface}
    \label{fig:gallery}
\end{figure}

\cref{fig:gallery} shows the interface of \MakeUppercase{gallery}.
\label{sec:gallery}

\subsection{\MakeUppercase{about} Interface}
\begin{figure}[htb]
    \centering
    \includegraphics[width=\columnwidth]{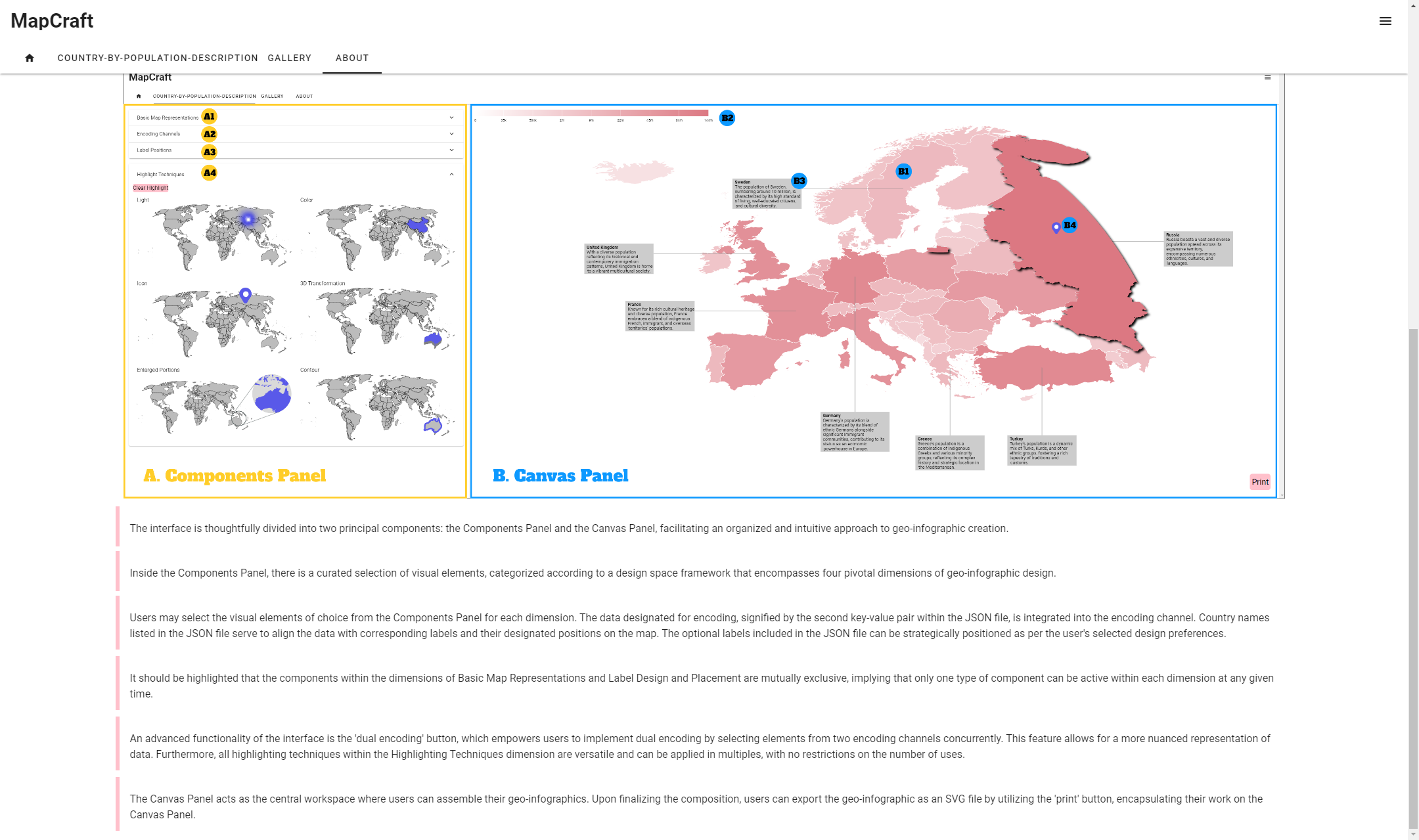}
    \caption{\MakeUppercase{about} Interface}
    \label{fig:about}
\end{figure}

\cref{fig:about} shows the interface of \MakeUppercase{about}.
\label{sec:about}

\section{Collected Geo-infographics Created by Users}
\begin{figure}[htb]
    \centering
    \includegraphics[scale=0.1]{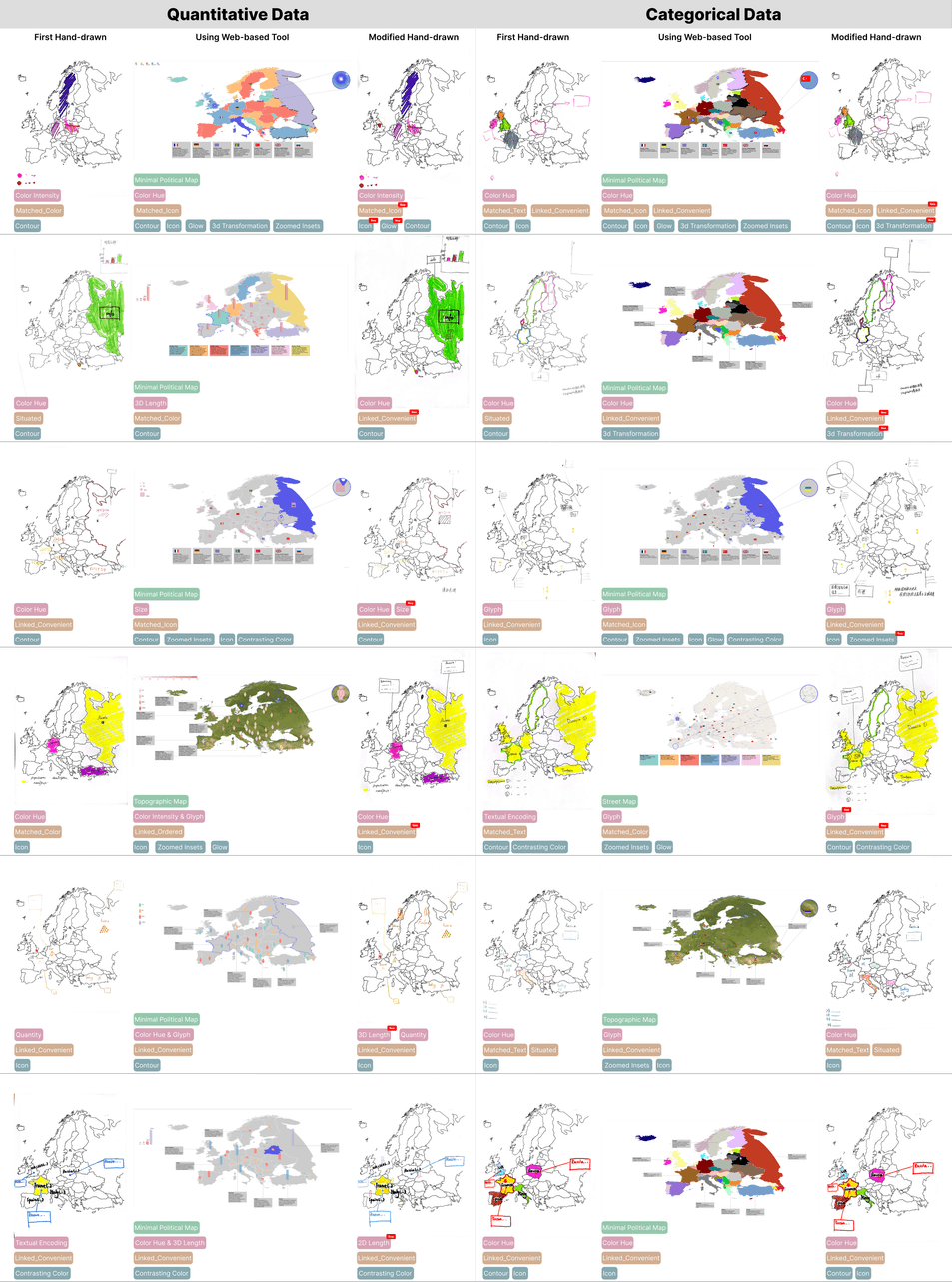}
    \caption{Geo-infographics Created by Users}
    \label{fig:usersGraphic1}
\end{figure}

\begin{figure}[htb]
    \centering
    \includegraphics[scale=0.1]{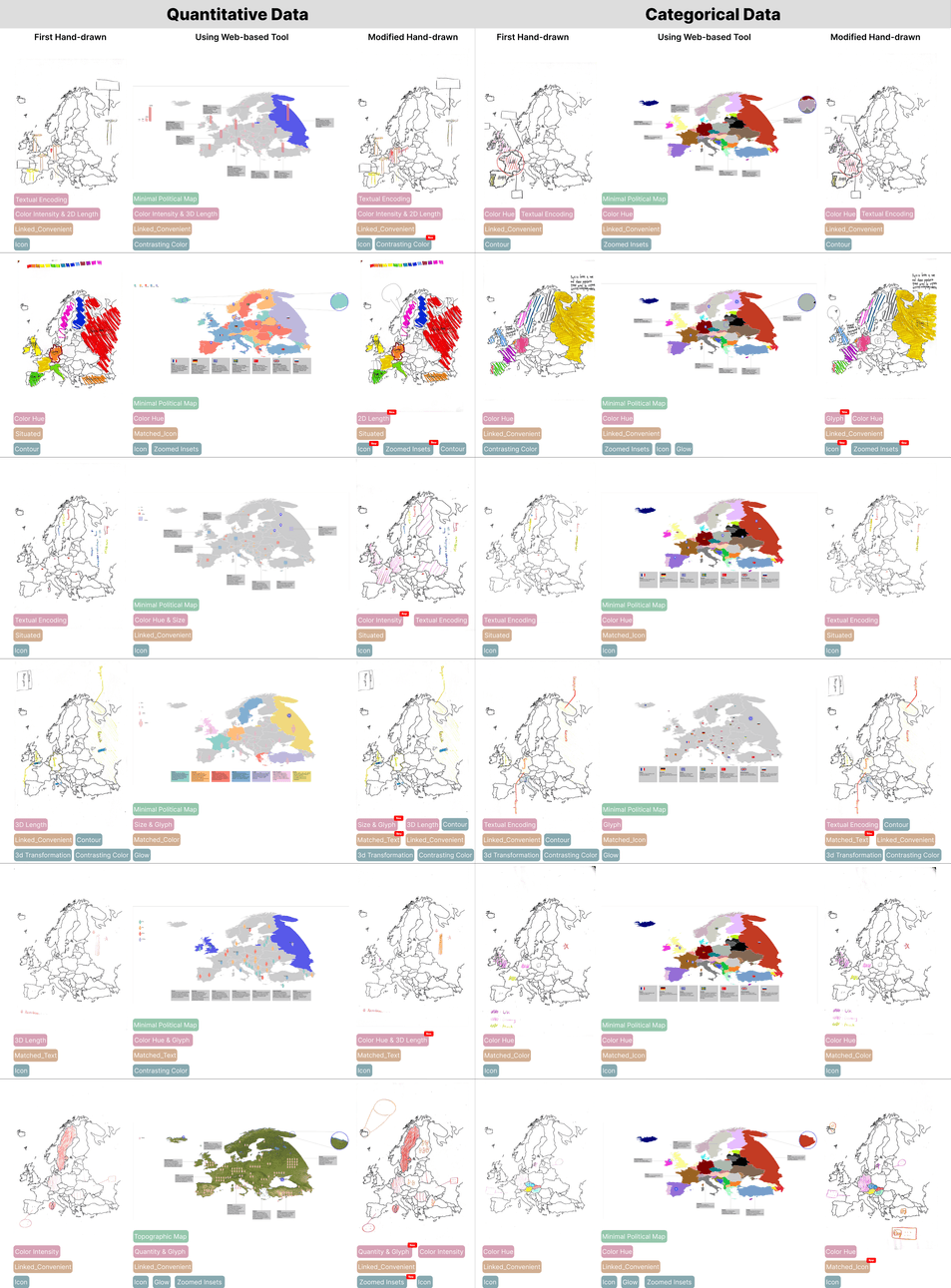}
    \caption{Geo-infographics Created by Users}
    \label{fig:usersGraphic2}
\end{figure}

\cref{fig:usersGraphic1} and \cref{fig:usersGraphic2} show the geo-infographics created by users.
\label{sec:usersGraphic}

\section{User Study Questionnaire}
\subsection{Demographic Background}
Here are the contents of the demographic background questionnaire in the pre-questionnaire, consisting of six short answer questions and a multiple choice question.
\begin{itemize}[left=0pt]
\item \noindent \textbf{Question 1: }
Age.
\item \noindent \textbf{Question 2: }
Gender 
\item \noindent \textbf {Question 3: }
Filed of Study or Profession.
\item \noindent \textbf {Question 4: }
Have you had any experience with data visualization?
\item \noindent \textbf {Question 5 (multiple choice question): }
Which of the following tools have you used for data visualization?
\begin{itemize}
    \item Microsoft Excel
    \item Tableau
    \item Google Charts
    \item D3.js
    \item R (ggplot2, Plotly)
    \item Python (Matplotlib, Seaborn, Plotly)
    \item Figma
    \item Other
\end{itemize}
\item \noindent \textbf{Question 6: }
Have you had any experience with Geo-infographics or interactive maps websites?
\item \noindent \textbf {Question 7: }
Self-assessment of confidence level in design geo-infographic.
\end{itemize}
\label{sec:background}

\subsection{Geo-infographics Designing Knowledge}
Here are the contents of the geo-infographics designing knowledge assessment in the pre- and post-questionnaire, consisting of eight short answer questions.
\begin{itemize}[left=0pt]
\item \noindent \textbf{Question 1: }
What encoding channels come to your mind for representing population data?
\item \noindent \textbf{Question 2: }
What encoding channels come to your mind for representing country information data?
\item \noindent \textbf {Question 3: }
Based on population data, which encoding channels do you think can be effectively combined for dual encoding?
\item \noindent \textbf {Question 4: }
Trying to add annotation in geo-infographic, what placement or match method can you think of?
\item \noindent \textbf {Question 5: }
What methods can you think of for highlighting within a geo-infographic?
\item \noindent \textbf{Question 6: }
Which encoding channels within the design space do you think are appropriate for population data?
\item \noindent \textbf {Question 7: }
Which encoding channels within the design space do you think are appropriate for country information data?
\item \noindent \textbf{Question 8: }
Which encoding channels within the design space do you think can be effectively combined for dual encoding based on population data?
\end{itemize}
\label{sec:knowledge}

\subsection{Interview}
Here are the questions in the interview.
\begin{itemize}[left=0pt]
\item \noindent \textbf{Question 1: }
What tool features are particularly useful for you to complete tasks? Please explain the reason.
\item \noindent \textbf{Question 2: }
What difficulties or challenges did you encounter while using \textit{MapCraft}?
\item \noindent \textbf {Question 3: }
Do you think it is reasonable to categorize visual elements into basic map representations, encoding channels, label design and placement, and highlighting techniques? Is it easy to understand? Has it caused you any trouble?
\item \noindent \textbf {Question 4: }
Do you plan to continue using this tool in future projects? Why or why not? What factors will influence your decision?
\end{itemize}
\label{sec:interview}

\end{document}